\documentclass[cernpreprint,USenglish]{na61doc}
\usepackage{lineno, blindtext}
\usepackage{graphicx}
\usepackage{hyperref}
\usepackage[T1]{fontenc}
\usepackage[utf8]{inputenc}
\usepackage{amsmath}
\usepackage{float}
\usepackage{enumerate}
\usepackage{tikz}
\usetikzlibrary{patterns}
\usepackage{bbding}
\usepackage{caption}
\usepackage{makecell}
\usepackage{varwidth}
\usepackage{flafter}

\newcommand{\GeV}{\ensuremath{\mbox{Ge\kern-0.1em V}}\xspace}
\newcommand{\GeVc}{\ensuremath{\mbox{Ge\kern-0.1em V}\!/\!c}\xspace}
\newcommand{\AGeV}{\ensuremath{A\,\mbox{Ge\kern-0.1em V}}\xspace}
\newcommand{\AGeVc}{\ensuremath{A\,\mbox{Ge\kern-0.1em V}\!/\!c}\xspace}
\newcommand{\pt}{$p_{\text{T}}$}

\newcommand{\coordinate}[1]{{\fontfamily{lmss}\selectfont#1}}

\newcommand{\GeantFour}{{\scshape Geant4}\xspace}

\newcommand{\EposLong}{{\scshape Epos1.99}\xspace}

\newcommand{\pp}{\mbox{\textit{p+p}}\xspace}

\definecolor{kPink+2}{RGB}{204,102,153}
\definecolor{kOrange+8}{RGB}{255,102,51}
\definecolor{kGreen+2}{RGB}{0,153,0}
\definecolor{kCyan+2}{RGB}{0,153,153}
\definecolor{kBlue+2}{RGB}{0,0,153}
\definecolor{kRed+1}{RGB}{204,0,0}
\definecolor{kBlue}{RGB}{0,0,204}
\definecolor{kBlue-9}{RGB}{153,153,255}
\definecolor{kGreen}{RGB}{0,153,0}
\definecolor{kRed}{RGB}{204,0,0}
\definecolor{kCyan}{RGB}{51,204,204}
\definecolor{kMagenta}{RGB}{153,0,153}
\definecolor{kPink}{RGB}{204,0,102}
\definecolor{kGray}{RGB}{204,204,204}
\definecolor{kBlack}{RGB}{0,0,0}

\definecolor{kRed+3}{RGB}{102,0,0}
\definecolor{kRed+2}{RGB}{153,0,0}
\definecolor{kRed-4}{RGB}{255,51,51}
\definecolor{kRed-7}{RGB}{255,102,102}
\definecolor{kRed-9}{RGB}{255,153,153}
\definecolor{colorPSDCentral}{RGB}{66,129,164}
\definecolor{colorPSD150}{RGB}{72,169,166}
\definecolor{colorPSD75}{RGB}{212,180,131}
\definecolor{colorPSD19}{RGB}{193,102,107}

\PreprintIdNumber{CERN-EP-2025-056}

\ShineJournal{}

\ShineJournalRef{}
\ShineDOI{}
\ShineTitle{Multiplicity and net-electric charge fluctuations in central Ar+Sc interactions at 
13$A$, 19$A$, 30$A$, 40$A$, 75$A$, and 150\AGeVc
beam momenta measured by \NASixtyOne at the CERN SPS}
\ShineAbstract{
This paper presents results on multiplicity fluctuations of positively and negatively charged hadrons as well as net-electric charge fluctuations measured in central Ar+Sc interactions at beam momenta 13$A$, 19$A$, 30$A$, 40$A$, 75$A$, and 150\AGeVc. The fluctuation analysis is one of the tools to search for the predicted critical point of strongly interacting matter. Results are corrected for the experimental biases and quantified using cumulant ratios. In most instances, multiplicity and net-charge distributions appear narrower than the corresponding Poisson or Skellam distributions. Cumulant ratios are compared with the \EposLong model predictions, which provide a qualitative description that aligns with observations for positively and negatively charged particles. 
The obtained results are also compared to earlier \NASixtyOne results from inelastic \pp interactions in the same analysis acceptance.\\
\begin{center}
    
\end{center}

}
\begin{document}

\maketitle

\section{Introduction}
This paper presents measurements of the multiplicity and net-electric charge event-by-event fluctuations in central ${}^{40}$Ar+${}^{45}$Sc collisions at beam momenta ($p_\mathrm{beam}$) of 13$A$, 19$A$, 30$A$, 40$A$, 75$A$, and 150\AGeVc by \NASixtyOne at the CERN Super Proton Synchrotron (SPS). The 
corresponding energy per nucleon pair in the center-of-mass system is $\sqrt{s_{\mathrm{NN}}}=5.1, 6.1, 7.6, 8.8, 11.9$, and 16.8 GeV. This study is a part of the \NASixtyOne strong interaction program \cite{Antoniou:2006mh, Abgrall:2014fa} and is devoted to the search for the critical point of strongly interacting matter~\cite{Fodor:2004nz}. The study of the phase diagram of strongly interacting matter is the 
main aim of the two-dimensional scan of the collision energy and the colliding system size performed by the \NASixtyOne experiment. 

The measurement of fluctuations in nucleus-nucleus ($A$+$A$) collisions is one of the most promising tools for searching for the critical point (CP) as fluctuations are sensitive to the change of the correlation length in the system~\cite{Stephanov:1999zu, Stephanov:2004xs}. The fluctuations are also sensitive to the size of the system and varying collision geometry. 
Appropriate statistical methods and careful data selection are employed to mitigate the impact of these unwanted effects on the presented results. In this paper, so-called intensive fluctuation measures are utilized for the comparison between different system sizes. Also, only central Ar+Sc collisions were selected for the analysis to reduce varying collision geometry. The presented intensive quantities were already successfully used; for details, see Refs.~\cite{Mackowiak-Pawlowska:2021tch, ALICE:2021hkc, STAR:2014egu}.

The paper is organized as follows. In Section~\ref{sec:fluctMeas}, intensive measures of fluctuations are briefly introduced. The experimental setup of \NASixtyOne is presented in Sec.~\ref{sec:detector}. Data processing and analysis procedures are described in Sec.~\ref{sec:analysis}. The results and their comparison with model calculations and \pp results are discussed in Sec.~\ref{sec:results}. Section~\ref{sec:summary} contains the summary of this study.\\

The following variables and definitions are used in this paper. The particle rapidity $y$ is calculated
assuming pion mass in the collision center-of-mass system (cms), $y=0.5\ln{[(E+p_{\text{L}})/(E-p_{\text{L}})]}$, where $E$
and $p_{\text{L}}$ are the particle energy and longitudinal momentum, respectively. The transverse component of the momentum is denoted as \pt, the electric charge as $q$, and the azimuthal angle as  $\phi$. It is defined as the angle between the transverse momentum vector and the horizontal (\coordinate{x}) axis. The collision energy per nucleon pair in the center-of-mass system is denoted as $\sqrt{s_{\text{NN}}}$.
\section{Measures of multiplicity and net-charge fluctuations}
\label{sec:fluctMeas}

The shape of the multiplicity and net-electric charge (for simplicity called further net-charge) distributions can provide information about the structure of the strongly interacting matter phase diagram~\cite{Stephanov:1999zu}. It is quantified by moments, cumulants, and a combination of these quantities~\cite{Asakawa:2015ybt}. The multiplicity and net-charge fluctuations may be sensitive to the presence of CP, and this sensitivity increases with the order of measured quantities~\cite{Vovchenko:2015pya, Asakawa:2015ybt}. 

Intensive quantities are the quantities independent of the volume of the system ($V$) within the ideal Boltzmann gas model described by the Grand Canonical Ensemble (IB-GCE)~\cite{Begun:2006jf}, or of the number of wounded nucleons ($W$) within the Wounded Nucleon Model ~\cite{Bialas:1976ed}. 
Intensive quantities can be calculated by dividing two extensive quantities, which are proportional to the size of the system. Extensive quantities are, for example, cumulants ($\kappa_i$, where $i$ is the order of the cumulant) or the algebraic or central moments of the distribution. For instance, the mean multiplicity of produced charged hadrons $\langle N\rangle$ is $\kappa_1$ and the variance of the distribution $\sigma^2 = \langle N^2\rangle - \langle N\rangle^2$ is $\kappa_2$. The third-order cumulant, $\kappa_{3}$, is equal to the third central moment $\mu_{3}=\langle(N-\langle N\rangle)^{3}\rangle$ and $\kappa_{4}$ is given by $\mu_{4}-3\sigma^{4}$, where $\mu_{4}=\langle(N-\langle N\rangle)^{4}\rangle$.

The intensive quantities studied in this paper are scaled variance, scaled skewness, and scaled kurtosis. If there are no fluctuations, the value of intensive quantities is equal to zero~\cite{Begun:2017gsw}. If the multiplicity varies according to the Poisson distribution, the value of intensive quantities equals unity. In IB-GCE, the net-charge, defined as a difference of positive and negative charges, is given by the Skellam distribution (not Poisson). For this reason, the intensive quantities of the net-charge distribution are slightly modified
to keep zero and unity as reference values. As a result, the final quantities considered in this paper are obtained as cumulant ratios in
the following way: 
\begin{enumerate}[(i)]
    \item for multiplicity distributions:
    \begin{align}
         \frac{\kappa_2}{\kappa_1}~, \quad  \frac{\kappa_3}{\kappa_2}~, \quad 
         \text{and}\quad \frac{\kappa_4}{\kappa_2}~.
    \end{align}
    \item for net-charge:
    \begin{align}
        \frac{\kappa_2}{\kappa_1^+ + \kappa_1^- }~, \quad \frac{\kappa_3}{\kappa_1}~, \quad \text{and} \quad \frac{\kappa_4}{\kappa_2}~,
    \end{align}
    where $\kappa_1^+$ and $\kappa_1^-$ are the first cumulants of multiplicity distribution for particles of the corresponding charge. 
\end{enumerate}

In nucleus-nucleus collisions, it is experimentally challenging to determine the volume of the created system precisely. Collisions of similar volume are grouped in so-called centrality classes, from peripheral to central ones. The influence of the volume fluctuations on the intensive quantities for systems of different sizes is discussed in Ref.~\cite{Mackowiak-Pawlowska:2021sea}. Volume fluctuations are addressed by selecting the most central collisions where the fluctuations are relatively small. Moreover, it is a known fact that the kinematic acceptance of the detector may affect measured ratios~\cite{Asakawa:2015ybt}. Thus, a well-defined phase-space region in rapidity, transverse momentum, and azimuthal angle is also provided. For details, see Sec.~\ref{sec:analysis}.
\section{\NASixtyOne detector}
\label{sec:detector}
\begin{figure*}[t]
  \centering
    \includegraphics[width=\textwidth]{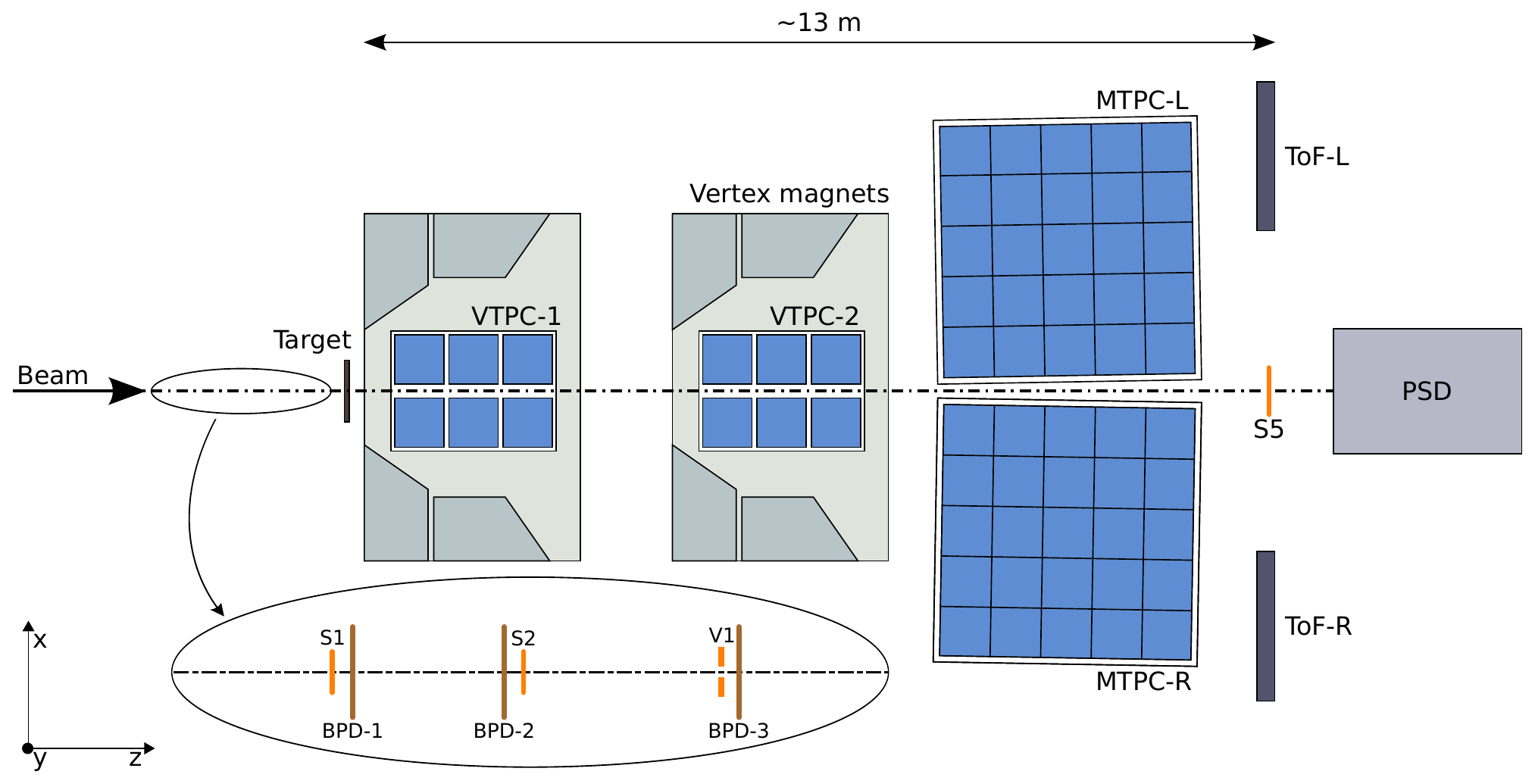}
  \caption{The schematic layout of the \NASixtyOne~\cite{Abgrall:2014fa} detector as used for the $^{40}$Ar+$^{45}$Sc energy scan (horizontal cut, not to scale). The inset provides a closer view of the beam and trigger detector configuration. The plot illustrates the alignment of the \NASixtyOne coordinate system, with its origin located in the middle of VTPC-2 along the beam axis. The \coordinate{z} axis represents the beam direction, while the magnetic field is designed to bend charged particle trajectories in the \coordinate{x}--\coordinate{z} (horizontal) plane. The drift direction in the TPCs aligns with the \coordinate{y} (vertical) axis.}
  \label{fig:setup}
\end{figure*}

The \NASixtyOne detector, depicted in Fig.~\ref{fig:setup}, is a large-acceptance hadron spectrometer situated in CERN North Area.
Beam delivery to the detector occurs via the H2 beamline from the Super Proton Synchrotron (SPS)~\cite{Abgrall:2014fa}. Three Beam Position Detectors (BPDs) and a set of scintillator counters (S1, S2, V1) are utilized upstream of the spectrometer to measure beam position and time references. 

The detection system consists of four Time Projection Chambers (TPCs) responsible for tracking charged particles produced in collisions. Positioned downstream of the target and along the beamline are two Vertex TPCs (VTPCs) enclosed within superconducting magnets with a combined bending power of up to 9~Tm. The magnetic field is scaled proportionally to the beam momentum to maintain momentum acceptance similar for different collision energies. Placed symmetrically on both sides of the beamline, downstream of the VTPCs, are the Main TPCs (MTPCs) and two walls of pixel Time-of-Flight (ToF-L/R) detectors. The VTPCs are filled with an Ar:CO$_{2}$ gas mixture in the ratio 90:10, while the MTPCs utilize a mixture of 95:5.

Downstream of the MTPCs, the Projectile Spectator Detector (PSD) is located. It is a high-resolution calorimeter centered on the beam. It measures the energy flow around the beam direction allowing for selecting the desired collision centrality. 

\subsection{Target}

For the collisions of $^{40}$Ar+$^{45}$Sc nuclei, the experiment utilized targets obtained from Stanford Advanced Materials. The targets were $2\times2$ cm plates with thicknesses of 2 mm and 4 mm, comprising more than 99\% of Sc~\cite{Banas:2018sak}. Both plates were used during the data collection. The target was positioned at \coordinate{z} $\approx-580$ cm (upstream of the VTPC-1 front wall) within a special target holder~\cite{Abgrall:2014fa}. This target holder maintained a helium atmosphere around the target, reducing the likelihood of off-target interactions that would result from the beam interacting with the air in the vicinity of the target.

\subsection{Beam and triggers}

The $^{40}$Ar beam delivered by the SPS accelerator was the primary beam explicitly designed for \NASixtyOne.
The experiment successfully conducted collisions using beams at six different momenta: 13\textit{A}, 19\textit{A}, 30\textit{A}, 40\textit{A}, 75\textit{A}, and 150\AGeVc with the scandium targets.

Two scintillation counters, S1 and S2, are responsible for defining the beam, along with a veto counter, V1, featuring a 1~cm diameter hole, defining the beam position before reaching the target. Counter S1 also serves as the timing reference (start time) for all counters. The trigger signal to detect beam particles requires a coincidence of these three signals:
\begin{equation}
    \textrm{T1} = \textrm{S1} \wedge\textrm{S2} \wedge\overline{\textrm{V1}}~. 
\end{equation}
The T4 trigger is based on the breakup of the beam ion due to interactions both in and downstream of the target. It indicates a minimum bias interaction within the target, necessitating an incoming beam particle signal (T1) and a signal lower than that of beam nuclei from the S5 counter. The S5 counter, a scintillator with a diameter of 2~cm, is positioned on the beam downstream of the MTPCs. The T4 trigger condition is denoted as:

\begin{equation}
\textrm{T4} = \textrm{T1} \wedge \overline{\textrm{S5}}~.
\end{equation}

Additionally, a threshold is set in sixteen central PSD modules to select approximately 20-30\% of the most central collisions (this value varies with different beam momenta). These modules mainly measure the energy of the projectile spectators, so a signal below the set threshold indicates central interaction. The trigger logic for events meeting these criteria is:

\begin{equation}
\textrm{T2} = \textrm{T4} \wedge \overline{\textrm{PSD}}~.
\end{equation}
The presented results are obtained using the events collected with the T2 trigger.
\section{Analysis}
\label{sec:analysis}
The objective of the analysis was to measure the fluctuations in multiplicity and net-charge of charged hadrons produced in strong interactions and electromagnetic decays in the 1\%~most central $^{40}$Ar+$^{45}$Sc collisions within the selected acceptance. The acceptance is defined by the $y - p_{\mathrm{T}} - \phi$ region, representing the high-efficiency area of the detector in the case of Ar+Sc interactions~\cite{Detector_acceptance}. The analysis procedure involved the following steps:
\begin{enumerate}[(i)]
    \item data selection based on the event and track selection criteria to ensure the highest data quality,
    \item evaluation of the distributions of charged hadron multiplicities and net-charge,
    \item correction of the charged hadron multiplicity and net-charge distributions using the unfolding method,
    \item calculation of the fluctuation measures from the corrected distributions,
    \item estimation of the statistical and systematic uncertainties of the results.
\end{enumerate}

\subsection{Data processing}
\label{subsec:dataProcess}

\subsubsection{Event selection}

\begin{figure}[h]
              \centering
              \includegraphics[width=0.5\textwidth]{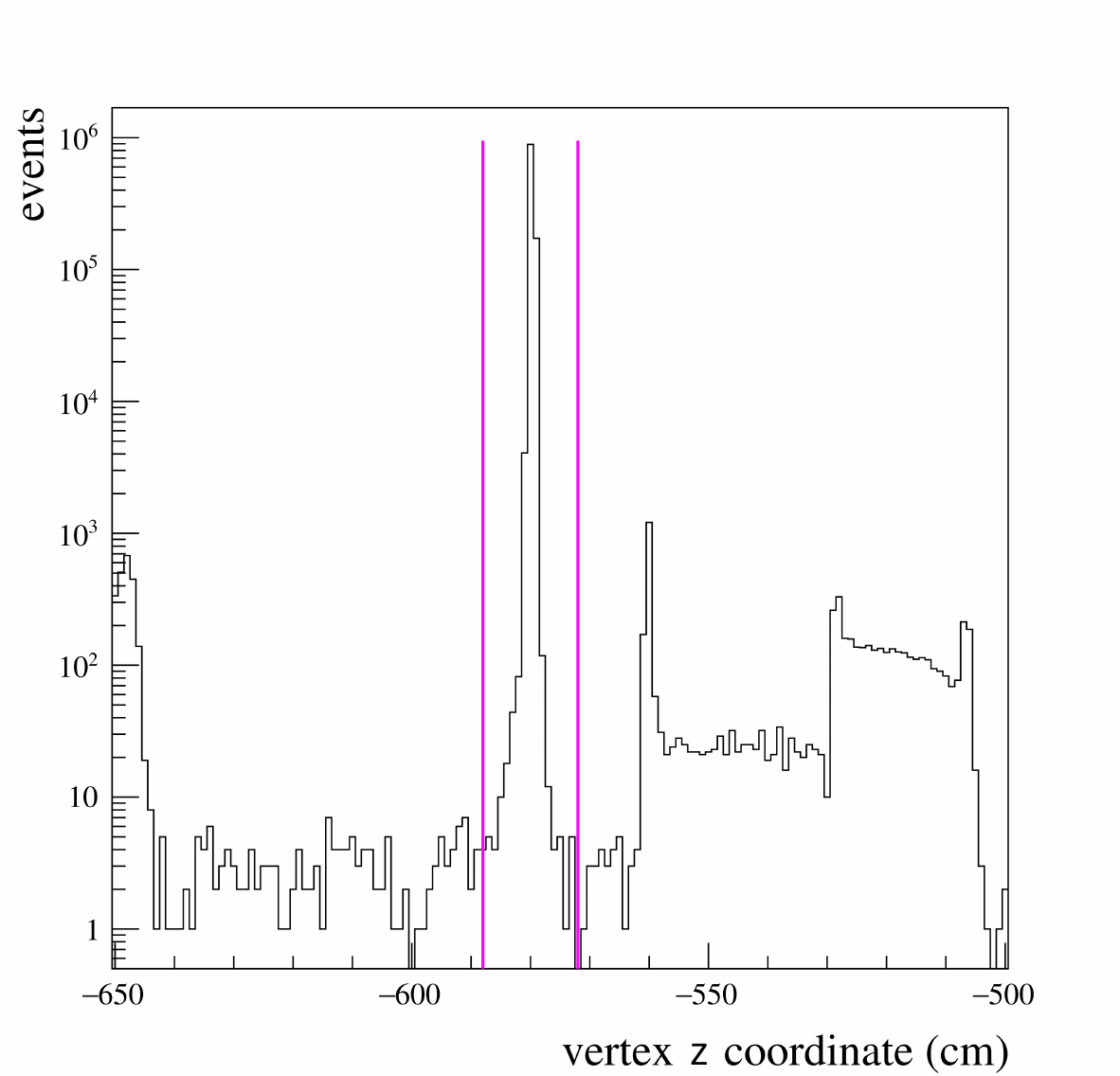}
              \caption{Distribution of the \coordinate{z} coordinate of the fitted primary vertex for the T2-triggered events of $^{40}$Ar+$^{45}$Sc interactions at 150\AGeVc. The vertical lines indicate the \coordinate{z} region selected for the analysis. The \coordinate{z} position of the center of the target is equal to $-$580 cm.}
              \label{fig:vtxZcut}
          \end{figure}

The analysis involved several selection criteria to ensure the quality and reliability of the data. These criteria are as follows:
\begin{enumerate}[(i)]
    \item The initial event preselection was carried out using the hardware central interaction trigger T2, specifically configured to use signals from the sixteen small PSD modules (see Sec.~\ref{sec:subsecCentrality}). T2 accepted a fraction of the most central interactions, characterized by the lowest forward energy deposited in these modules. The fraction depends on the specific beam momentum.\label{evetCon1}
    \item Within the analysis, events with any off-time beam particles within a time window of $\pm$4~$\mu$s around the trigger particle were excluded. Furthermore, events with any interaction detected within a time window of $\pm$25~$\mu$s around the trigger particle were also eliminated.\label{evetCon2}
    \item For an accurate measurement of the beam particle trajectory, it was essential to have simultaneous signals in either BPD-1 or BPD-2, along with the signal in BPD-3. \label{evetCon3}
    \item The main vertex \coordinate{z}-coordinate of the event has to be between $\pm 8$ cm around the center of the Sc target (see Fig. \ref{fig:vtxZcut})\label{evetCon4},
    \item An upper cut on the energy $E_{\text{PSD}}$ measured in the PSD in order 
    to select the 1\% most central collisions was introduced. For the detailed centrality selection procedure, see Sec.~\ref{sec:subsecCentrality}.
\end{enumerate}
The number of events remaining for the analysis is listed in Table~\ref{tab:data_details}. 

\begin{table}[h]
    \centering
    \begin{tabular}{p{1.2cm}|p{1.2cm}|p{2.7cm}}
        $p_{\mathrm{beam}}$ (\GeVc) & $\sqrt{s_{\mathrm{NN}}}$ (GeV) & 1\%~most central events\\    
        \hline
        13$A$   & 5.1 & 49435\\
        19$A$   & 6.1 & 52408\\
        30$A$   & 7.6 & 91019\\
        40$A$   & 8.8 & 129751\\
        75$A$   & 11.9 & 116103\\
        150$A$  & 16.8 & 46452\\
    \end{tabular}
    \caption{Numbers of accepted events that passed the event selection criteria.}
    \label{tab:data_details}
\end{table}

\subsubsection{Track selection}

To focus on the primary charged hadron distributions and minimize contamination from secondary and off-time interactions, as well as weak decays, the following track selection criteria were implemented:

\begin {enumerate}[(i)]
    \item the track momentum fit at the primary vertex should have converged,
    \item the total number of reconstructed points on the TPC track should be greater than 30,
    \item the sum of the number of reconstructed points in VTPC-1 and VTPC-2 should be greater than~15,
    \item the distance between the track extrapolated to the interaction plane and the vertex (track impact parameter)
    is required to be smaller than 4~cm in the horizontal (bending) plane and 2~cm in the vertical (drift) plane,
    \item the mean ionization energy loss measured for a given track does not indicate an electron or positron candidate,
    \item a track remains in the high-efficiency region of the detector~\cite{Detector_acceptance}, see Fig.~\ref{fig:acceptance}.
    \begin{figure}[h]
        \centering
        \includegraphics[width=0.49\textwidth]{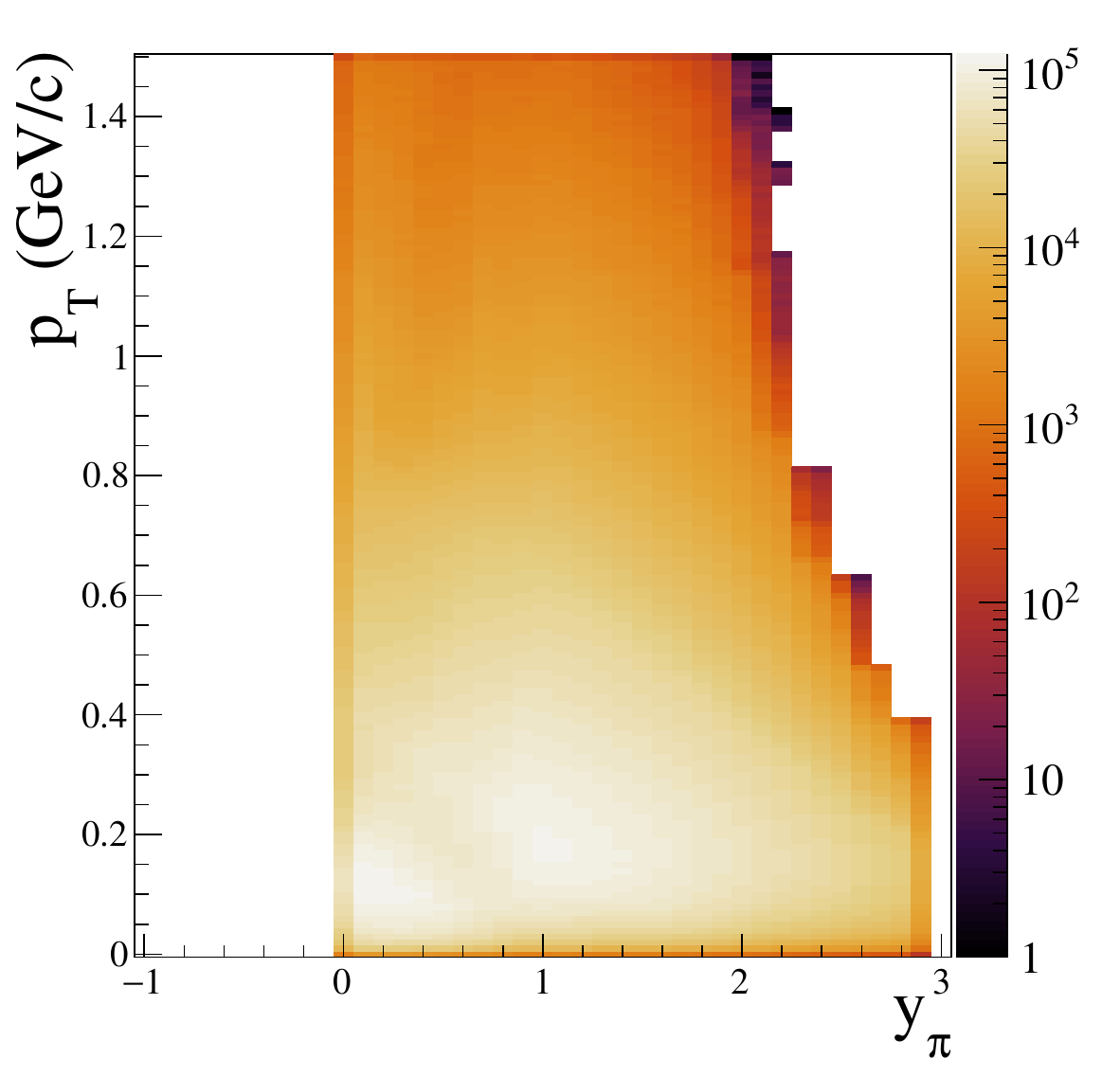}
        \includegraphics[width=0.49\textwidth]{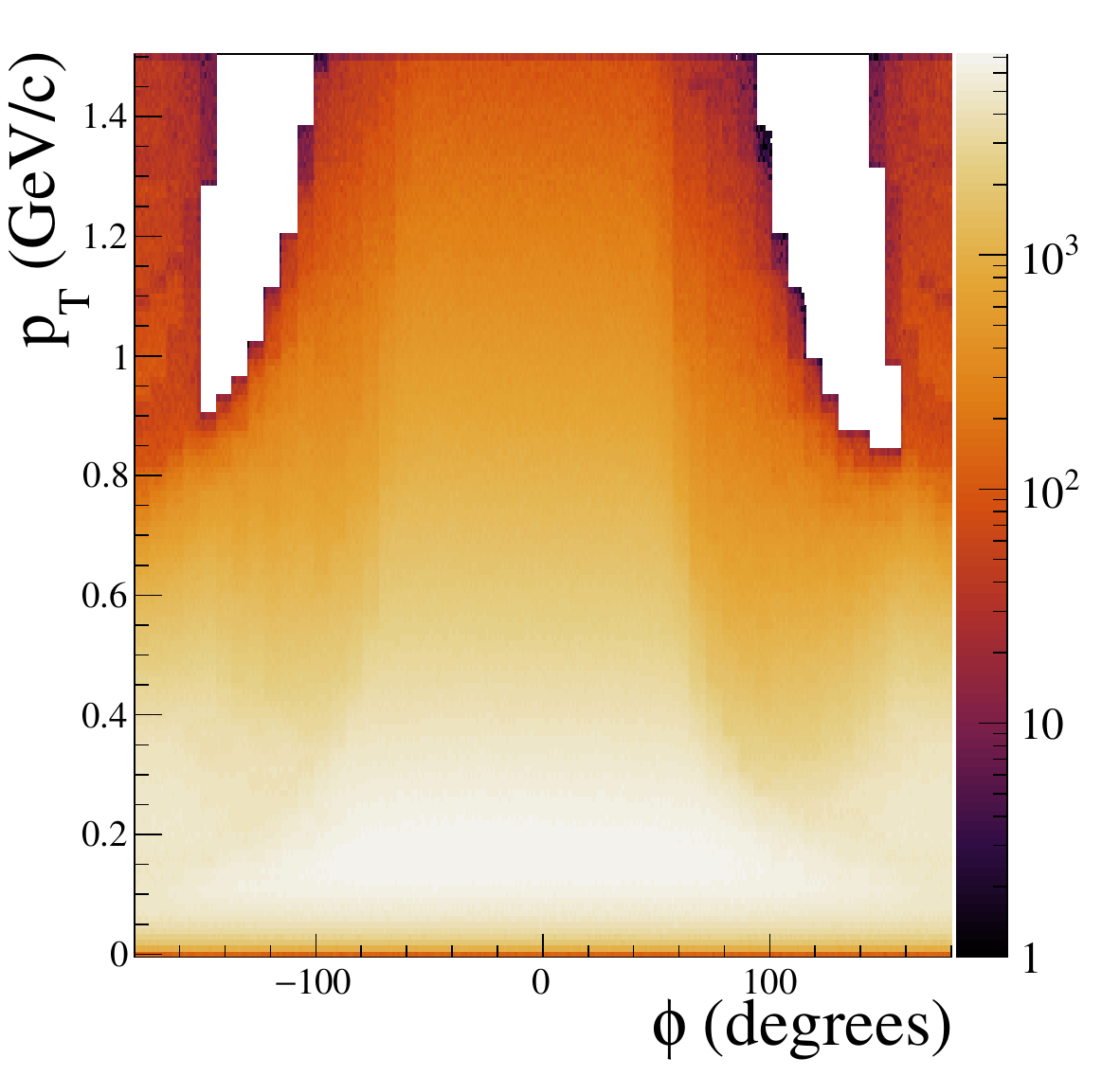}
        \caption{
        Distribution of charged particle tracks in the acceptance region selected for the analysis in the $y - p_{\mathrm{T}} - \phi$ space, for $^{40}$Ar+$^{45}$Sc interactions at 150\AGeVc. The rapidity $y_{\pi}$ of each particle is calculated assuming pion mass.
        }
        \label{fig:acceptance}
    \end{figure}
\end {enumerate}

\subsubsection{Selection of central collisions}
\label{sec:subsecCentrality}

This section presents the selection procedure of the 1$\%$ most central events using the PSD calorimeter. 
The final results presented in this paper refer to the $1\%$ of Ar+Sc collisions with the lowest value of the
forward energy $E_\mathrm{F}$ (most central collisions).
The quantity $E_{\mathrm{F}}$ represents the total energy deposit of all particles produced in the $^{40}$Ar+$^{45}$Sc collision via strong and electromagnetic processes within the forward rapidity region defined by the PSD acceptance map \cite{PSD_acceptance}, as well as nuclear fragments emitted in this region. While $E_\mathrm{F}$ can be obtained from models, it is not directly available in the experimental data. 
\begin{figure}[h]
       \centering 
       \begin{minipage}[m]{0.45\textwidth}
       	\centering 
       \includegraphics[width=0.6\textwidth]{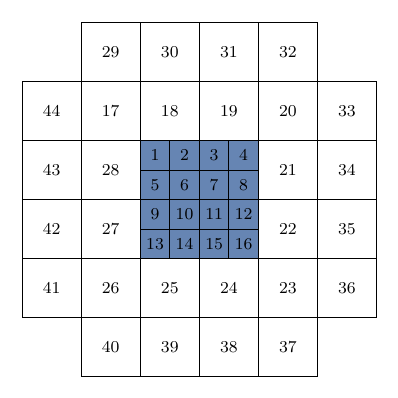}\\
       T2 trigger
   		\end{minipage}
       \begin{minipage}[m]{0.45\textwidth}
       	\centering 
       \includegraphics[width=0.6\textwidth]{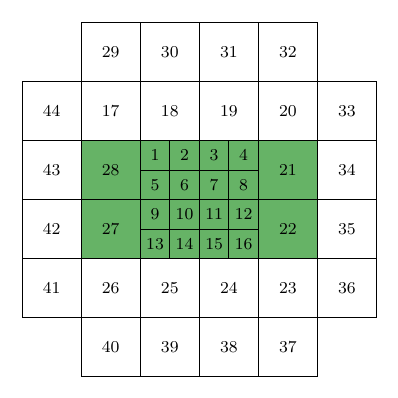}\\
       150\AGeVc
   		\end{minipage}\\\vspace{0.5cm}
       \begin{minipage}[m]{0.45\textwidth}
       	\centering 
       \includegraphics[width=0.6\textwidth]{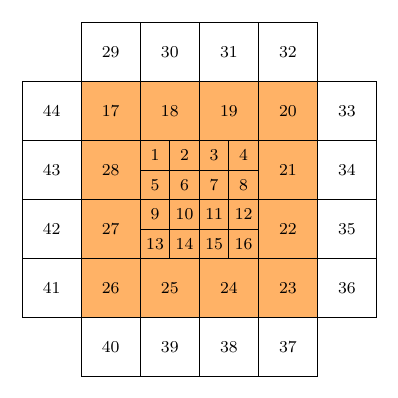}\\
       75\textit{A}, 40\textit{A}, 30\AGeVc
   		\end{minipage}
       \begin{minipage}[m]{0.45\textwidth}
       	\centering 
       \includegraphics[width=0.6\textwidth]{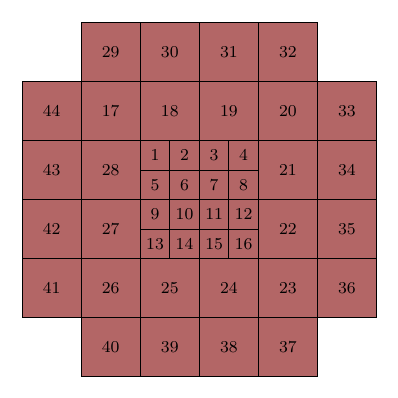}\\
       19\textit{A}, 13\AGeVc
       \end{minipage}
       \caption{Schematic diagrams indicating by shading the PSD modules used in the online and offline event selection. The trigger is derived from the energy in the central 16 modules (1-16) in blue color. Determination of the PSD energy $E_\text{PSD}$ uses the green (150\AGeVc), orange (75\textit{A}, 40\textit{A}, 30\AGeVc) or all modules (19\textit{A}, 13\AGeVc) at the respective beam momenta.}

    \label{fig:psd_selection}
\end{figure}
Instead, the \NASixtyOne experiment measures $E_\mathrm{PSD}$ in PSD which corresponds to $E_\mathrm{F}$. A specific subset of PSD modules is used for its calculation to ensure the closest possible approximation of $E_{\mathrm{PSD}}$ to $E_\mathrm{F}$. The precise value of $E_\mathrm{PSD}$, its correspondence to $E_\mathrm{F}$ and the percentage of selected centrality were based on performed tests~\cite{Mackowiak-Pawlowska:2021sea, Seryakov:2018kzj}. Figure~\ref{fig:psd_selection} shows which modules were used to obtain $E_{\mathrm{PSD}}$ at a given beam momentum.
The selection of these modules is optimized for each beam momentum by analyzing the correlation between the mean module energy and the total multiplicity of charged particles registered in the TPCs. The $E_\mathrm{PSD}$ calculation is performed on the data preselected by the hardware central interaction trigger T2.
The example distribution of $E_{\mathrm{PSD}}$ is shown in Fig.~\ref{fig:PSD_dist}. For details on T2 selection see Ref.~\cite{NA61SHINE:2021nye}. \\
In Monte Carlo simulations, centrality was selected directly using either $E_\mathrm{F}$ or the simulated value of $E_{\mathrm{PSD}}$. The simulated $E_{\mathrm{PSD}}$ represents energy stored in the simulated PSD.
The simulation accounts for various PSD effects on energy measurements, including energy leakage, energy smearing, and measurement resolution. Its purpose is to replicate the impact of the PSD on $E_\mathrm{F}$ accurately. Both $E_\mathrm{F}$ and simulated $E_\mathrm{PSD}$ play crucial roles in the correction procedure, allowing for the differentiation of various event classes.

\begin{figure}[h]
    \centering
    \includegraphics[width=0.49\textwidth]{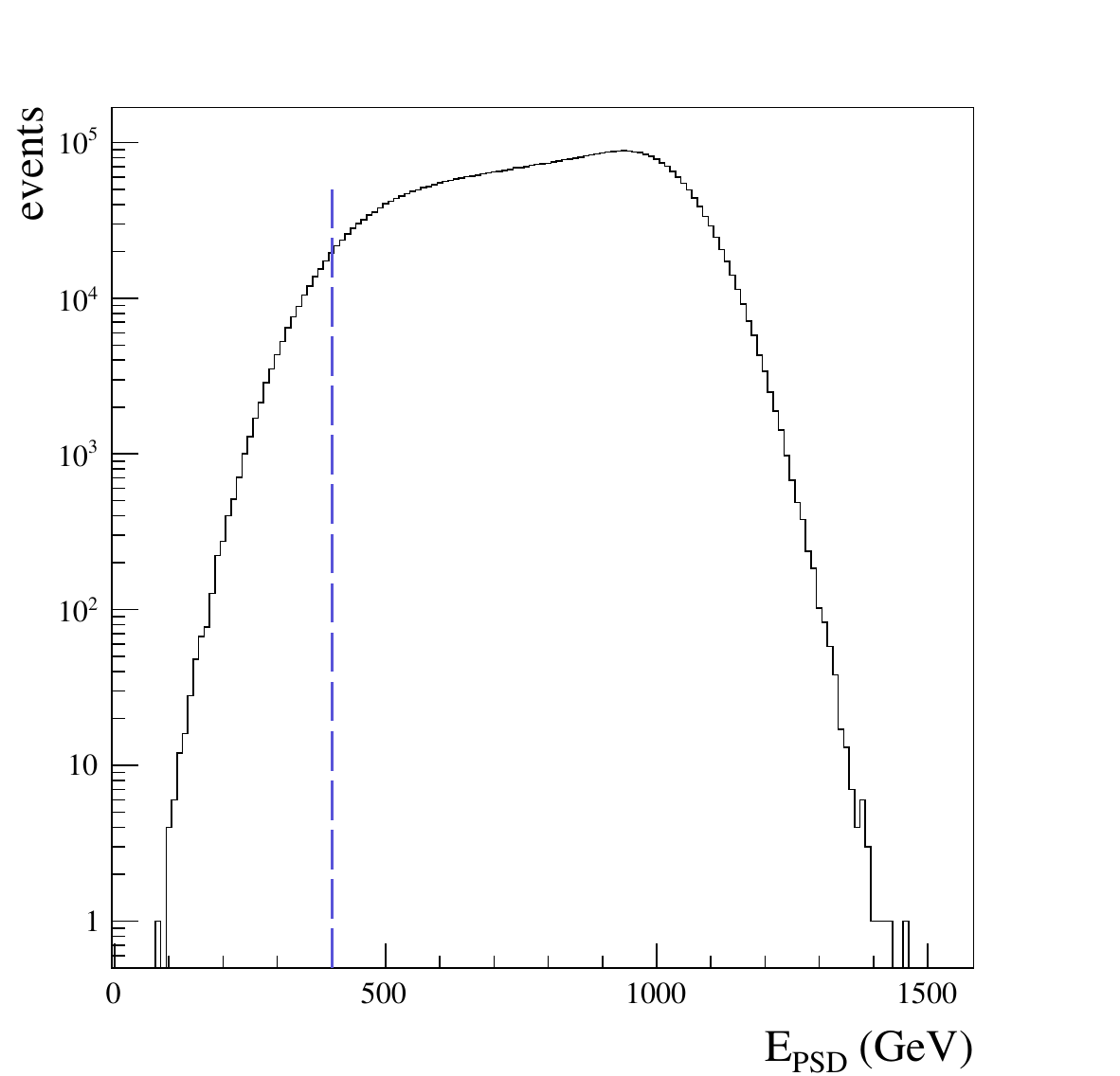}
    \includegraphics[width=0.49\textwidth]{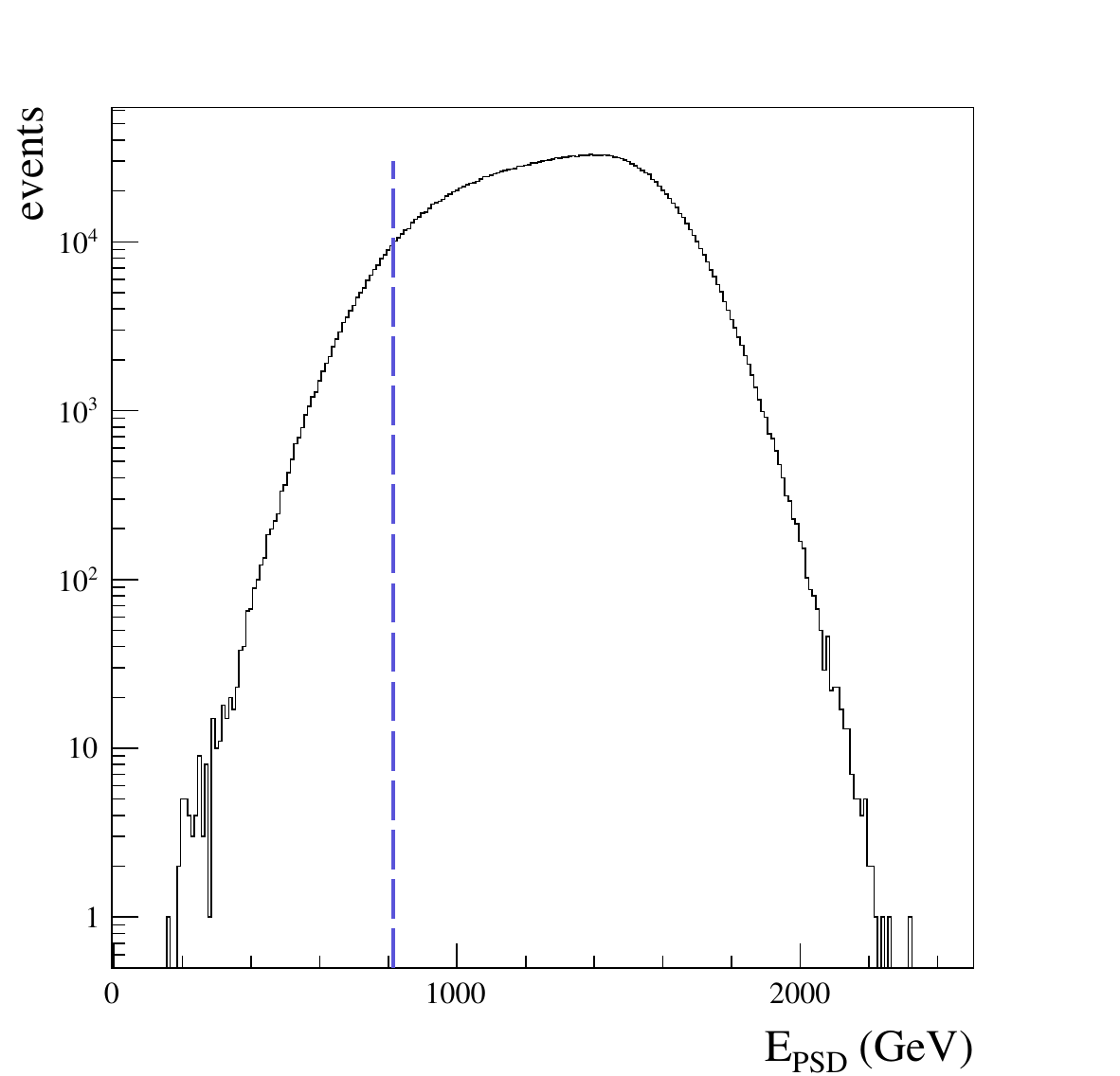}
    \caption{Event centrality selection using the energy $E_\mathrm{PSD}$ measured by the PSD calorimeter (for selection of modules see Fig.~\ref{fig:psd_selection}). Distributions show measured $E_\mathrm{PSD}$ for the T2 selected events for $40A$ (\textit{left}) and $75$\AGeVc (\textit{right}) beam momenta. The left part of the distribution marked with the vertical line is the region with the $1\%$ most central events.}
    \label{fig:PSD_dist}
\end{figure}

\subsection{Corrections}
The data correction procedure addresses four main biases of this analysis:
\begin{enumerate}[(i)]
\item Gain of peripheral events: this refers to cases where $E_\mathrm{F}$ of non-central events is incorrectly reconstructed, leading to them being mistakenly classified as central events;
\item Loss of central events: imperfectness of event reconstruction or presence of off-time interaction/beam particle may lead to the loss of central events; 
\item Gain of tracks: imperfectness of the track selection leading to accepting tracks coming from other sources, e.g. gain of tracks coming from weak decays; 
\item Loss of tracks: the reconstruction efficiency may lead to the loss of some tracks that should have been included in the analysis.
\end{enumerate}
The influence of off-target interactions is negligible as only very central collisions are considered, and a helium target holder surrounds the target (see Sec.~\ref{sec:detector}). 

To address all four biases, a one-dimensional unfolding based on Iterative Bayes' theorem \cite{DAgostini} was chosen. For this purpose, the open-access framework RooUnfold \cite{unfolding} was utilized. The one-dimensional unfolding process involves using a response matrix built from Monte Carlo simulations. This matrix allows for the correction of the data histogram to account for the effects of the detector response. To perform the unfolding, a regularization parameter must be specified. It depends on the specific algorithm selected for the unfolding. 
The choice of this parameter is made in such a way to minimize systematic uncertainties while preserving small statistical uncertainties. In the context of Bayesian unfolding, the regularization parameter corresponds to the number of iterations in the unfolding process. For each beam momentum, the optimal number of iterations in the unfolding procedure was determined through a systematic study. The fluctuation measures were calculated for the unfolded distributions using different numbers of iterations. The final number of iterations was selected such that 
a further increase in the number of iterations do not cause a significant change in the values of the fluctuation measures. The unfolding procedure involves categorizing events into three distinct types: "good", "missed", and "fake" events. These categories may have varying definitions depending on the specific experiments and analysis processes. In this particular study, the definitions are as follows:
\begin{enumerate}[(i)]
    \item \textbf{Good Events}: encompass all inelastic events that successfully pass the event selection criteria, meaning they are properly reconstructed in the TPCs. Additionally, these events must fall within a predefined centrality bin as determined by both $E_\mathrm{F}$ and simulated $E_\mathrm{PSD}$.
    
    \item \textbf{Missed Events}: this category includes simulated events that were not successfully reconstructed as they did not meet the event selection criteria. It also includes cases where events are reconstructed and fall within the predefined centrality bin based on $E_\mathrm{F}$ but have different centrality when assessed using the simulated $E_\mathrm{PSD}$.
    
    \item \textbf{Fake Events}: consist of correctly reconstructed events that fall within the predefined centrality bin determined by the simulated $E_\mathrm{PSD}$ but display a different centrality when evaluated using $E_\mathrm{F}$.
    
\end{enumerate}
The unfolding procedure relies on sufficient statistics of simulated events and a Monte Carlo model that encompasses the data distribution to yield meaningful results. It is crucial that the distribution of reconstructed Monte Carlo events is close to the entire data distribution. Otherwise, the part of the data that is not represented in Monte Carlo may lead to distortions due to the lack of information in the simulated data. To meet this requirement, a dedicated study was conducted at each beam momentum to determine the centrality of the simulated events, ensuring that reconstructed Monte Carlo multiplicities cover the data. 
Thus, one should remember that the response matrix is not necessarily built on the 1$\%$ most central interactions in simulated data but rather on the centrality bin, which covers all scenarios represented in the data. 
Depending on beam momentum, it varies between 10$\%$, 15$\%$, and 20$\%$ most central Ar+Sc collision simulated in \EposLong~\cite{Werner:2008zza}. Positively ($h^{+}$) and negatively ($h^{-}$) charged hadron distributions are unfolded with the means of 1D unfolding. Net-charge ($h^{+}-h^{-}$) can be unfolded in two ways. Either with 1D unfolding at the level of the final distribution or with the 2D unfolding of $h^{+}$ and $h^{-}$ distribution. Finally, 1D unfolding was used directly on the net-charge distribution, applying the same 0-1\% centrality selection for both the measured and Monte Carlo data. The effectiveness of this method was verified by comparing 1D and 2D unfolding results on a smaller system (\pp reactions), confirming that the difference between the two approaches was negligible \cite{NA61SHINE:2023kod}.

\begin{figure}
    \centering
    \includegraphics[width=0.325\textwidth]{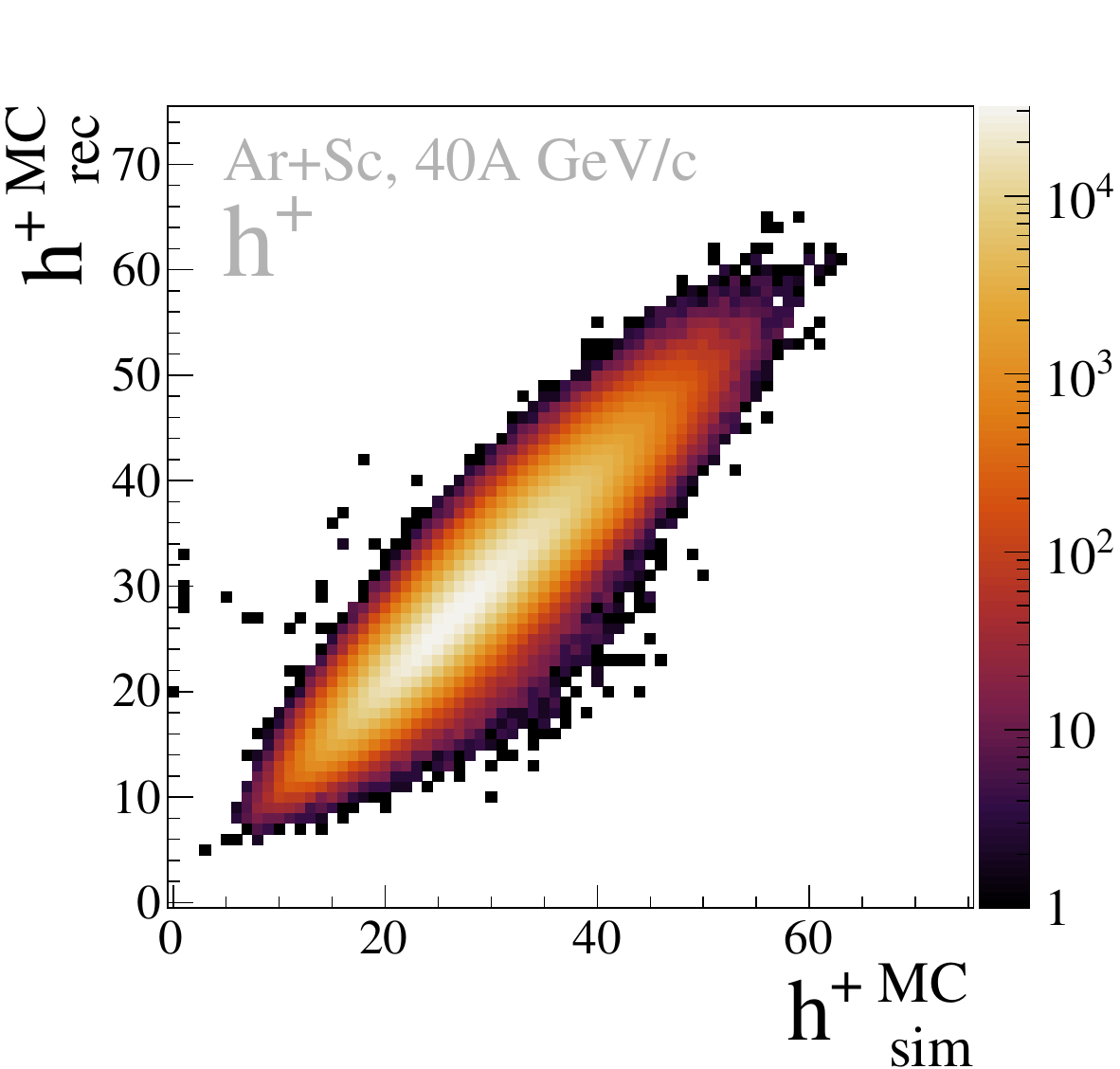}
    \includegraphics[width=0.325\textwidth]{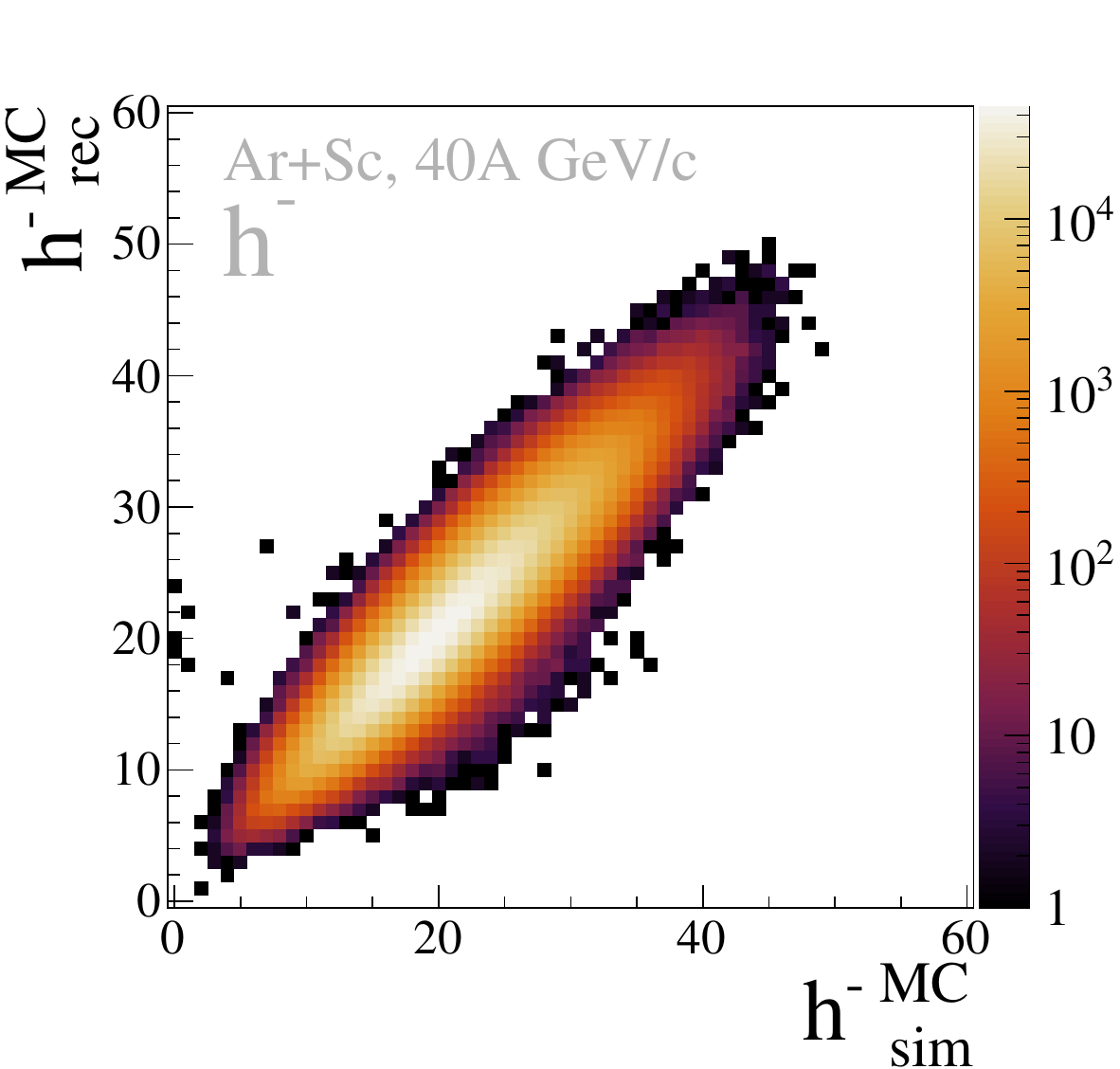}
    \includegraphics[width=0.325\textwidth]{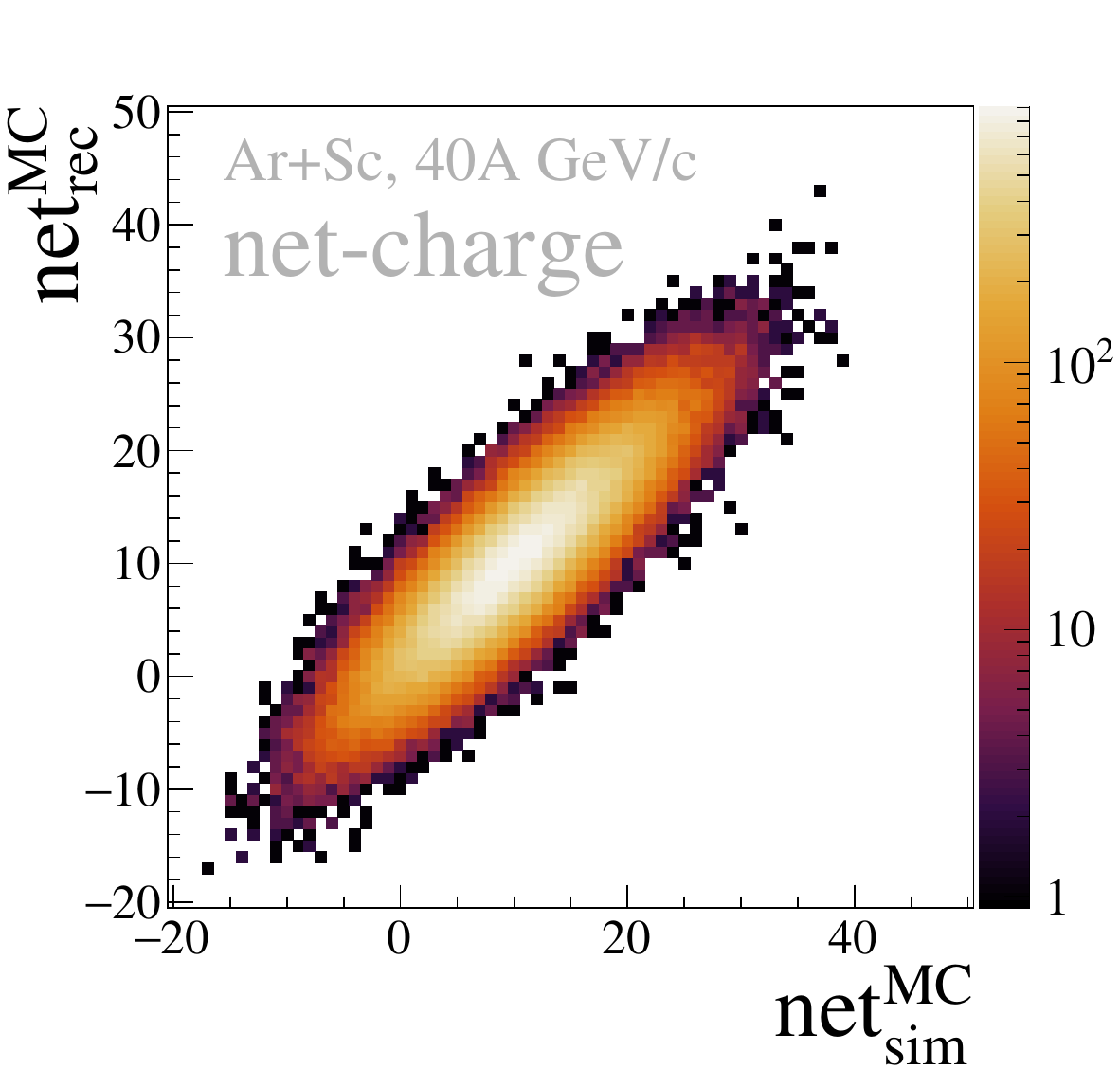}
    \caption{
    Relation between Monte Carlo (MC) simulated (sim) and reconstructed (rec) multiplicities for positively (\textit{left}) and negatively (\textit{middle}) charged particles as well as for net-charge (\textit{right}) in $^{40}$Ar+$^{45}$Sc interactions at beam momentum 40\AGeVc. These multiplicities enter the response matrix used in the unfolding procedure. Note that the different number of entries between $h^{+}$, $h^{-}$, and net-charge arises due to different centrality selections used for the construction of the response matrix.}
    \label{fig:respMtx}
\end{figure}

The Monte Carlo model selected as the generator for primary ${}^{40}$Ar+${}^{45}$Sc interactions is \EposLong. The simulation of particle propagation through the detector, including decays and secondary interactions, and the detector response were carried out in the \GeantFour environment~\cite{Agostinelli:2002hh}. This allowed for the generation of simulated events that closely resembled the actual experimental conditions. The simulated events were then processed through the standard \NASixtyOne reconstruction chain, where the detector signals were reconstructed and tracks were identified. 
The reconstructed and simulated Monte Carlo multiplicities and net-charge used in the response matrix are shown in Fig.~\ref{fig:respMtx}. Measured data distributions, as well as corrected ones, are shown in Fig.~\ref{fig:multiplicityDistrCorr}.

\subsection{Statistical uncertainties}

The bootstrap method~\cite{moore} was employed in this study to estimate statistical uncertainties. This approach estimates correlated uncertainties without fragmenting the initial data sample into smaller subsets – a division that would further deplete statistical precision. 
The core principle underlying the {bootstrap method} involves generating $M$ samples (referred to as {bootstrap samples}) from the original data set through sampling with replacement. In this process, each entry from the data set is eligible for inclusion in the {bootstrap sample} multiple times. The size of each bootstrap sample (representing the distribution of multiplicity or net-charge) corresponds to the size of the original data set. The number of bootstrap iterations was set to $M=500$. The final uncertainty of a given intensive quantity corresponds to the standard deviation of the distribution of this quantity derived from $M$ samples. The statistical uncertainties ($\sigma_\mathrm{stat}$) are presented in Table~\ref{tab:numResults}. 

\subsection{Systematic uncertainties}

The following effects were included in the estimation of the systematic uncertainty of the results of intensive fluctuation measures presented in this paper:
\begin{enumerate}[(i)]
    \item The uncertainty arising from the PSD resolution, and consequently the uncertainty associated with centrality selection, was evaluated by manually distorting a measured signal in each PSD module using an empirically obtained formula~\cite{Cybowska:2023PhD}. The corresponding uncertainty was computed by comparing the intensive fluctuation measures derived using the actual PSD energy distribution to those obtained using a modified version of the PSD energy distribution and denoted as $\sigma_\mathrm{PSD}$. Notably, the contribution of this effect to the overall systematic uncertainty remains within the bounds of the statistical uncertainty of the results.\label{sigmaPSD}
    \item The removal of the events with off-time beam particles close in time to the trigger particle. The systematic uncertainty of this effect was calculated by changing the time window from $\pm 4$~$\mu$s to $\pm 5$~$\mu$s.\label{sigmaTL1}
    \item The uncertainty from the reconstructed vertex \coordinate{z}-position cut (used in the same form in the data and the Monte Carlo reconstructed events) was estimated by changing the width of the cut from $\pm 5$ cm to $\pm 9$ cm.\label{sigmaTL2}
    \item The uncertainty from the track selection was estimated by removing the impact parameter cut and changing the minimum number of reconstructed points to 10 in all TPCs and 10 in the VTPCs.\label{sigmaTL3}
\end{enumerate}
In the case of effects (\ref{sigmaTL1})-(\ref{sigmaTL3}) as they are interdependent, the largest contribution was selected, and denoted as $\sigma_\mathrm{tl}$. The total systematic uncertainties are calculated as:
\begin{equation}
    \sigma_\mathrm{sys}=\sqrt{\sigma_\mathrm{PSD}^{2}+\sigma_\mathrm{tl}^{2}},
\end{equation}
where $\sigma_\mathrm{PSD}$ refers to PSD resolution and $\sigma_\mathrm{tl}$ refers to combined effect of remaining effects. 
The total systematic uncertainties are presented in Table~\ref{tab:numResults}. 
\section{Results}
\label{sec:results}

\begin{figure}
    \centering
    \includegraphics[width=\textwidth]{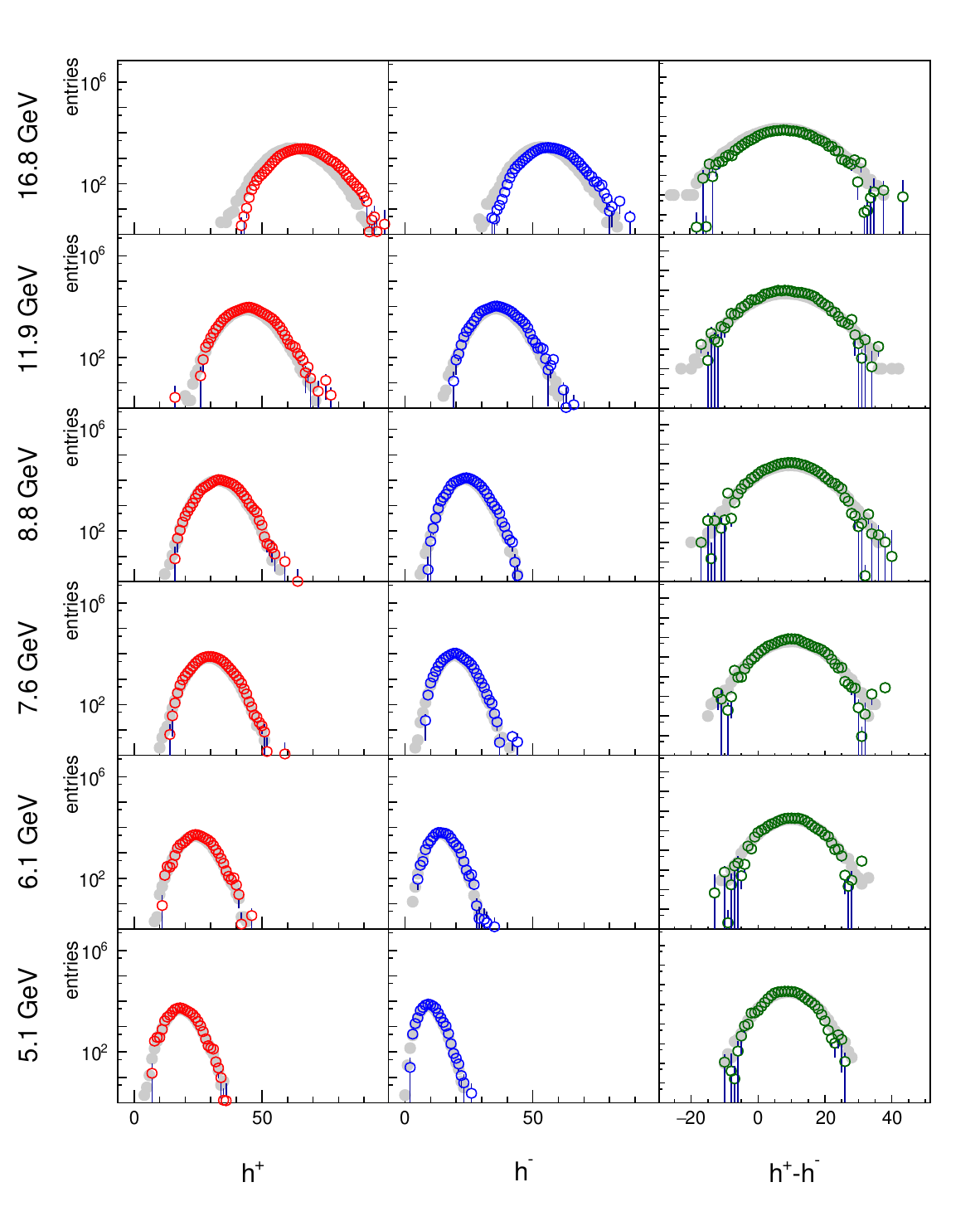}
    \caption{Positively (\textit{left}) and negatively (\textit{middle}) charged hadron multiplicity distributions, as well as net-charge (\textit{right}), for the 1\% most central $^{45}$Ar+$^{40}$Sc collisions at energies ${\sqrt{s_\mathrm{NN}} = 5.1, 6.1, 7.6, 8.8, 11.9, 16.8}$ GeV. The open circles stand for corrected distributions, while the gray, full circles stand for uncorrected ones.}
    \label{fig:multiplicityDistrCorr}
\end{figure}

The corrected and uncorrected multiplicity distributions of positively and negatively charged hadrons, as well as net-charge distributions in the 1$\%$ most central $^{40}$Ar+$^{45}$Sc interactions, are presented in Fig.~\ref{fig:multiplicityDistrCorr}.

\begin{figure}
    \centering
    \includegraphics[width=0.83\textwidth]{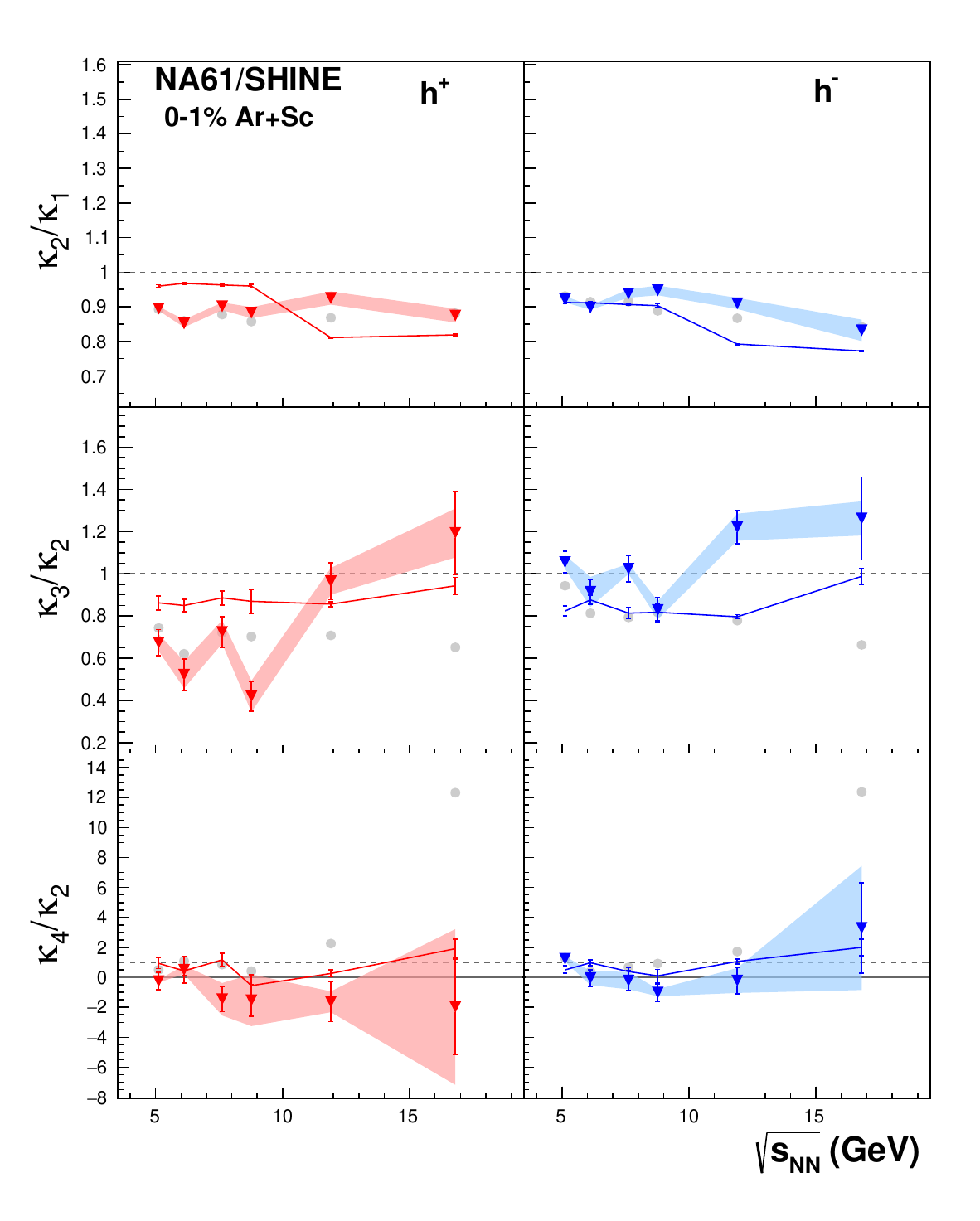}
    \caption{The energy dependence of $\kappa_2/\kappa_1$ (\textit{top}), $\kappa_3/\kappa_2$ (\textit{middle}), and $\kappa_4/\kappa_2$ (\textit{bottom}) for positively (\textit{left}) and negatively (\textit{right}) charged hadron multiplicity distribution in the 1$\%$ most central $^{40}$Ar+$^{45}$Sc interactions. Color triangles correspond to quantities obtained from corrected distributions, while gray circles correspond to those obtained from uncorrected distributions (statistical uncertainties not indicated). The error bars correspond to statistical uncertainties, while the color bands correspond to systematic uncertainties. The solid color lines show the \EposLong model predictions. The dashed line at unity corresponds to the reference value of the Poisson distribution. The solid line at zero corresponds to the case with no fluctuations in the system.}
    \label{fig:PosNegQuant}
\end{figure}

The energy dependence of corrected and uncorrected intensive quantities $\kappa_2/\kappa_1$, $\kappa_3/\kappa_2$, and $\kappa_4/\kappa_2$ for the multiplicity distribution of positively and negatively charged hadrons in the 1$\%$ most central $^{40}$Ar+$^{45}$Sc interactions is depicted in Fig.~\ref{fig:PosNegQuant}. Deviations from the reference value of unity indicate that obtained distributions deviate from Poisson distribution. The $\kappa_2/\kappa_1$ ratio of $h^{+}$ does not indicate any explicit non-monotonic behavior within the extent of systematic uncertainty. For both charges, $\kappa_{2}/\kappa_{1}$ is lower than one, indicating that the distribution is narrower than the reference. 
The ratio $\kappa_3/\kappa_2$ indicates a considerable difference between $h^{+}$ and $h^{-}$ at lower collision energies. In the case of positively charged hadrons, $\kappa_3/\kappa_2$ remains well below one (mostly visible for $\sqrt{s_{\mathrm{NN}}}=8.8$~\GeV). At higher collision energies, it starts to increase up to approximately one. It should be underlined that uncorrected data do not show such changes. The unfolding correction tends to remove peripheral events (see Fig.~\ref{fig:multiplicityDistrCorr}) that are wrongly recognized as central, thus reducing the low multiplicity tail of the distribution. This correction is energy-dependent and hence different for each energy, with the largest effect taking at top energy. In the case of negatively charged hadrons, one can not see such a large deviation from unity. At lower collision energies, $\kappa_3/\kappa_2$ of $h^{-}$ is close to one and then increases at higher energies. 
In the case of $\kappa_4/\kappa_2$ of positively charged hadrons, the non-monotonic behavior is absent, yet most energies exhibit a negative sign. On the contrary, the $\kappa_4/\kappa_2$ ratio of negatively charged hadrons stays at or close to zero. The substantial statistical and systematic uncertainties make conclusive interpretations regarding this behavior challenging.

\begin{figure}
    \centering
    \includegraphics[width=0.50\textwidth]{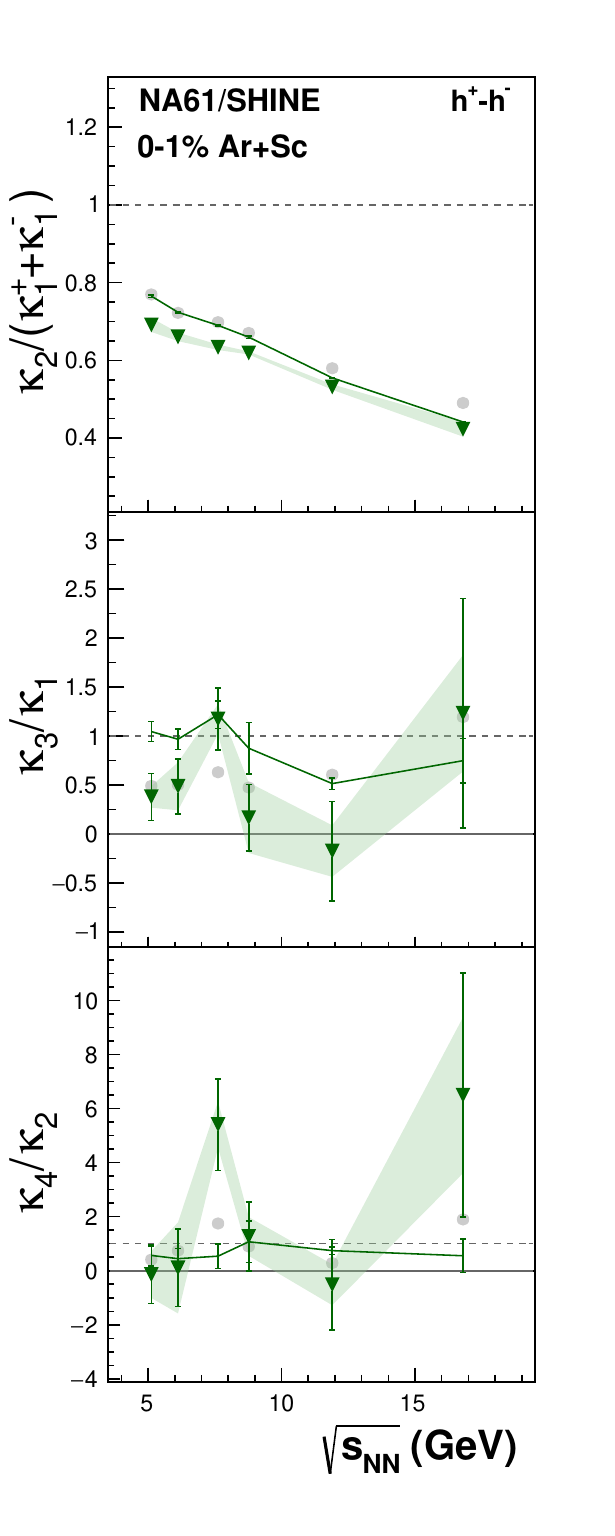}
    \caption{The energy dependence of $\kappa_2/(\kappa_1^+ + \kappa_1^-)$ (\textit{top}), $\kappa_3/\kappa_1$ (\textit{middle}), and $\kappa_4/\kappa_2$ (\textit{bottom}) of net-charge distirbution in the 1$\%$ most central $^{40}$Ar+$^{45}$Sc interactions. Color triangles correspond to quantities obtained from corrected distributions, while gray circles correspond to those obtained from uncorrected distributions (statistical uncertainties not indicated). The error bars correspond to statistical uncertainties, while the color bands correspond to systematic uncertainties. The solid color lines show the \EposLong model predictions. The dashed line at unity corresponds to the reference value of the Skellam distribution. The solid line at zero corresponds to the case with no fluctuations in the system.}
    \label{fig:NetQuant}
\end{figure}
Figure~\ref{fig:NetQuant} displays the energy dependence of the following intensive quantities: $\kappa_2/(\kappa_1^+ +\kappa_1^-)$, $\kappa_3/\kappa_1$, $\kappa_4/\kappa_2$, calculated for the net-charge distributions in the 1$\%$ most central $^{40}$Ar+$^{45}$Sc interactions. The $\kappa_2/(\kappa_1^+ +\kappa_1^-)$ (which is the net-charge counterpart of $\kappa_2/\kappa_1$ for multiplicity distributions) decreases with the increasing collision energy. 
It remains below the reference value without changing its sign, which indicates that the measured distribution is narrower than Skellam distribution. 
The ratios $\kappa_3/\kappa_1$ and $\kappa_4/\kappa_2$ display a hint of non-monotonic behavior with a maximum at $\sqrt{s_{\mathrm{NN}}}$=7.6 GeV, but one should keep in mind considerable associated uncertainties.

The \EposLong predictions for positively and negatively charged hadron multiplicity and net-charge fluctuations in Figs.~\ref{fig:PosNegQuant} and \ref{fig:NetQuant} are represented by the solid color lines. When examining the ratios of $\kappa_2/\kappa_1$, \EposLong tends to slightly overestimate $h^{+}$ at lower collision energies but underestimates them at higher collision energies. However, \EposLong reasonably describes $h^{-}$ at lower collision energies and underestimates them at higher collision energies.
In the case of $\kappa_{3}/\kappa_{2}$ of $h^{+}$ \EposLong does not predict energy dependence which is visible in the data. It stays relatively constant, predicting higher quantity values at lower energy and then lower at higher energies. For negative charge it remains below data signal for the considered energy range (except $\sqrt{s_{\mathrm{NN}}}=8.8$ GeV).

Regarding net-charge, 
\EposLong is of the same order of magnitude as the experimental results for all considered quantities.

The numerical values of quantities presented in Figs.~\ref{fig:PosNegQuant} and~\ref{fig:NetQuant} are listed in Table~\ref{tab:numResults}.

\begin{table}[h!]
    \centering
    \begin{tabular}{|c|c|c|c|c|c|c|c|}
    \hline
	beam momentum ($A$ GeV/\textit{c})	&13	&19	&30	&40	&75	&150	&\\
	\hline
$\kappa_2[h^+]$/$\kappa_1[h^+]$ &0.895&	0.852&	0.901&	0.883&	0.925&	0.875	&result\\
	&0.007&	0.007&	0.006&	0.005&	0.007&	0.013	&$\sigma_\mathrm{stat}$\\
&0.006&	0.011&	0.011&	0.016&	0.019&	0.020	&$\sigma_\mathrm{sys}$\\	
    \hline
	$\kappa_3[h^+]$/$\kappa_2[h^+]$ &0.674&0.522&0.724&0.419&0.964&1.194	&result\\    
	&0.061&	0.075&	0.073&	0.070&	0.087&	0.196	&$\sigma_\mathrm{stat}$\\
    &0.040&	0.064&	0.055&	0.074&	0.063&	0.116	&$\sigma_\mathrm{sys}$\\
    \hline
    $\kappa_4[h^+]$/$\kappa_2[h^+]$ &-0.251&0.504&-1.463&-1.524&-1.632&-1.976 &result\\
	&0.573&	0.893&	0.836&	1.061&	1.321&	3.161 &$\sigma_\mathrm{stat}$\\
    &0.092&	0.290&	1.083&	1.730&	0.695&	5.198 &$\sigma_\mathrm{sys}$\\
	\Xhline{4\arrayrulewidth}
	$\kappa_2[h^-]$/$\kappa_1[h^-]$	&0.921&0.898&0.938&0.947&0.910&0.832	&result\\
	&0.009&	0.008&	0.007&	0.006&	0.008&	0.013	&$\sigma_\mathrm{stat}$\\
 &0.009&	0.006&	0.012&	0.014&	0.016&	0.031	&$\sigma_\mathrm{sys}$\\
	
    \hline
	$\kappa_3[h^-]$/$\kappa_2[h^-]$	&1.055&0.914&1.024&0.828&1.221&1.262 &result\\
	&0.051&	0.059&	0.062&	0.059&	0.078&	0.197	&$\sigma_\mathrm{stat}$\\
 &0.032&	0.071&	0.024&	0.043&	0.064&	0.081	&$\sigma_\mathrm{sys}$\\
	
	\hline
$\kappa_4[h^-]$/$\kappa_2[h^-]$ &1.231&-0.055&-0.217&-1.023&-0.209&3.299	&result\\
	&0.450&	0.557&	0.655&	0.593&	0.893&	3.008 &$\sigma_\mathrm{stat}$\\
&0.185&	0.468&	0.581&	0.237&	0.832&	4.145	&$\sigma_\mathrm{sys}$\\
	
	\Xhline{4\arrayrulewidth}
$\kappa_2[h^{+}-h^{-}]/(\kappa_1[h^+]+\kappa_1[h^-])$ &0.691&0.660&0.633&0.619&0.531&0.422	&result\\
	&0.007&	0.007&	0.005&	0.004&	0.004&	0.006 &$\sigma_\mathrm{stat}$\\
&0.018&	0.011&	0.007&	0.004&	0.007&	0.019 &$\sigma_\mathrm{sys}$\\
	
    \hline
$\kappa_3[h^{+}-h^{-}]/\kappa_1[h^{+}-h^{-}]$ &0.380&0.487&1.174&0.167&-0.174&1.233	&result\\
	&0.240&	0.280&	0.317&	0.339&	0.508&	1.172	&$\sigma_\mathrm{stat}$\\
&0.110&	0.245&	0.152&	0.360&	0.263&	0.595	&$\sigma_\mathrm{sys}$\\
	
        \hline
 $\kappa_4[h^{+}-h^{-}]$/$\kappa_2[h^{+}-h^{-}]$ &-0.142&0.105&5.406&1.267&-0.520&6.491	&result\\
	&1.067&	1.436&	1.692&	1.285&	1.669& 4.514	&$\sigma_\mathrm{stat}$\\
 &0.864&	1.686&	0.854&	0.718&	0.757&	2.884	&$\sigma_\mathrm{sys}$\\
	
	\hline
    \end{tabular}
    \caption{Numerical values of $\kappa_2/\kappa_1$, $\kappa_3/\kappa_2$, $\kappa_4/\kappa_2$ for positively and negatively charged hadrons as well as $\kappa_2[h^{+}-h^{-}]/(\kappa_1[h^+]+\kappa_1[h^-])$, $\kappa_3[h^{+}-h^{-}]/\kappa_1[h^{+}-h^{-}]$, $\kappa_{4}[h^{+}-h^{-}]/\kappa_{2}[h^{+}-h^{-}]$ with their statistical and systematic uncertainties, in the 1$\%$ most central $^{40}$Ar+$^{45}$Sc interactions at 13$A$, 19$A$, 30$A$, 40$A$, 75$A$, and 150\AGeVc.}
    \label{tab:numResults}
\end{table}

\begin{figure}
    \centering
    \includegraphics[width=0.83\textwidth]{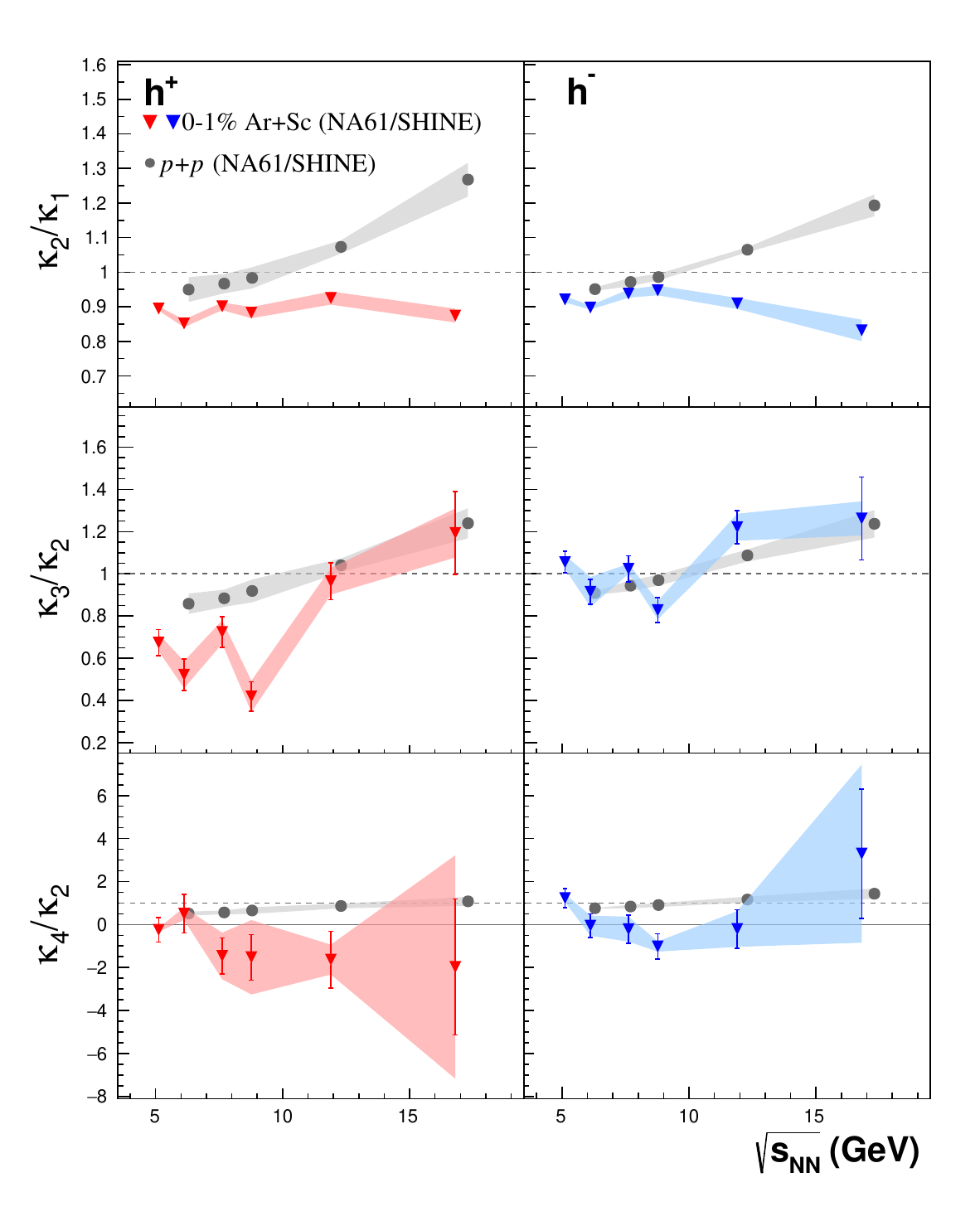}
    \caption{The energy dependence of $\kappa_2/\kappa_1$ (\textit{top}), $\kappa_3/\kappa_2$ (\textit{middle}), and $\kappa_4/\kappa_2$ (\textit{bottom}) for positively (\textit{left}) and negatively (\textit{right}) charged hadron multiplicity distribution in the 1$\%$ most central $^{40}$Ar+$^{45}$Sc (triangles) and inelastic \pp interactions (gray circles)~\cite{NA61SHINE:2023kod}. The error bars correspond to statistical uncertainties, while the color bands correspond to systematic uncertainties. The dashed line at unity corresponds to the reference value of the Poisson distribution. The solid line at zero corresponds to the case with no fluctuations in the system.}
    \label{fig:PosNegQuantUni}
\end{figure}
\begin{figure}
    \centering
    \includegraphics[width=0.50\textwidth]{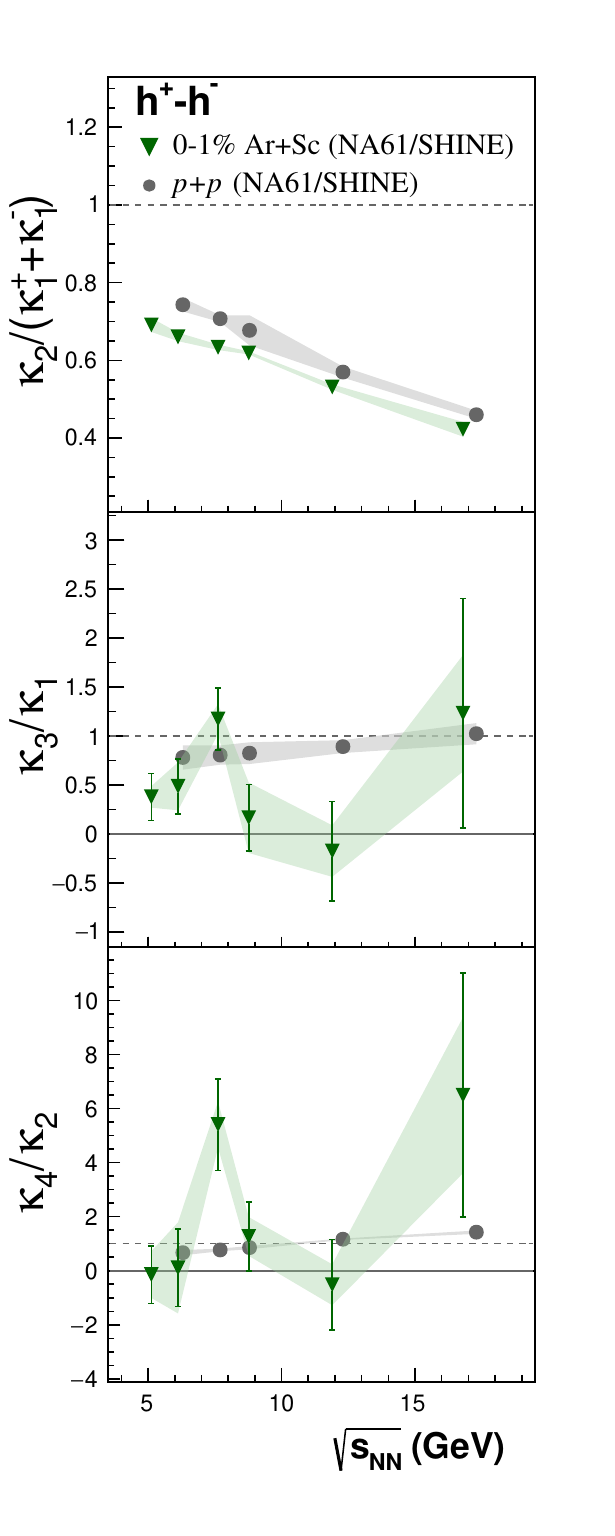}
    \caption{The energy dependence of $\kappa_2/(\kappa_1^+ + \kappa_1^-)$ (\textit{top}), $\kappa_3/\kappa_1$ (\textit{middle}), and $\kappa_4/\kappa_2$ (\textit{bottom}) of net-charge in the 1$\%$ most central $^{40}$Ar+$^{45}$Sc (triangles) and inelastic \pp interactions (gray circles)~\cite{NA61SHINE:2023kod}. The error bars correspond to statistical uncertainties, while the color bands correspond to systematic uncertainties. The dashed line at unity corresponds to the reference value of the Skellam distribution. The solid line at zero corresponds to the case with no fluctuations in the system.}
    \label{fig:NetQuantUni}
\end{figure}
The intensive quantities of charged hadron multiplicities in the 1$\%$ most central $^{40}$Ar+$^{45}$Sc collisions are compared with corresponding quantities in inelastic \pp~\cite{NA61SHINE:2023kod} collisions in Figs.~\ref{fig:PosNegQuantUni} and~\ref{fig:NetQuantUni}.
The trend is distinct in the case of $\kappa_2/\kappa_1$ for both positively and negatively charged hadron multiplicity. In \pp interactions, this quantity exhibits a consistent monotonic increase with collision energy. On the other hand, in the 1$\%$ most central $^{40}$Ar+$^{45}$Sc interactions, it remains roughly constant or tends to decrease with increasing energy. 
Furthermore, $\kappa_2/\kappa_1$ in \pp interactions surpasses the reference value of unity at higher collision energies, while in the 1$\%$ most central $^{40}$Ar+$^{45}$Sc interactions, it is below one. 
The increase of $\kappa_3/\kappa_2$ in both positively and negatively charged hadron multiplicity for $^{40}$Ar+$^{45}$Sc interactions follows a pattern similar to that observed in \pp collisions. It increases with collision energy but remains not far from the reference value. 
Notably, a systematic shift towards lower values is observed in $^{40}$Ar+$^{45}$Sc compared to \pp for positively charged hadrons at lower collision energies. 
In the case of $\kappa_4/\kappa_2$, the behavior diverges between $^{40}$Ar+$^{45}$Sc and \pp interactions. For the majority of collected energies $\kappa_4/\kappa_2$ in $^{40}$Ar+$^{45}$Sc collisions is below or close to \pp measurements indicating none or weak energy dependence. In contrast, \pp interactions show a slight increase with energy for $\kappa_4/\kappa_2$ of both charge types. It should be underlined that the substantial uncertainties in the $^{40}$Ar+$^{45}$Sc results complicate the precise assessment of these differences.

The signal magnitudes of $\kappa_{2}/(\kappa^{+}_{1}+\kappa^{-}_{1})$, $\kappa_{3}/\kappa_{1}$, and $\kappa_{4}/\kappa_{2}$ of net-charge in $^{40}$Ar+$^{45}$Sc interactions within estimated uncertainties are comparable with the signal measured in \pp interactions. In the case of $\kappa_{3}/\kappa_{1}$ large statistical uncertainties prevent a more detailed comparison of reactions. Taking into account uncertainties the largest deviation between reactions for $\kappa_{4}/\kappa_{2}$ can be observed at $\sqrt{s_{\mathrm{NN}}}=7.6$ GeV. Further studies in Ar+Sc interactions require either increasing the data statistics or finding a more suitable set of quantities that can be utilized in larger centrality bins where volume fluctuations can not be neglected.
\section{Summary and conclusions}
\label{sec:summary}

The primary objective of this study was to explore the critical behavior through an examination of fluctuations in multiplicity of positively and negatively charged hadrons, as well as net-charge, within the 1\%~most central collisions of $^{40}$Ar+$^{45}$Sc reactions at various energies ($\sqrt{s_\mathrm{NN}}$ = 5.1, 6.1, 7.6, 8.8, 11.9, and 16.8~GeV). 
The experimental results refer to cumulant ratios of multiplicity and net-charge distributions of primary charged hadrons resulting from strong interactions and electromagnetic decays in the 1\% most central $^{40}$Ar+$^{45}$Sc collisions within the selected acceptance. They were obtained by correcting measured data using the unfolding technique. 

The resulting ratios of cumulants deviate from the Poisson/Skellam distribution indicated by unity for the considered quantities differently than in the case of \pp reactions (except for net-charge $\kappa_{2}/(\kappa_{1}^{+}+\kappa_{1}^{-}$)). In most instances, multiplicity and net-charge distributions appear narrower than the corresponding Poisson or Skellam distributions. 
The magnitudes of the measured signal in the 1$\%$ most central $^{40}$Ar+$^{45}$Sc collisions and in \pp are comparable. Only in the case of $\kappa_{2}/\kappa_{1}$ of positively and negatively charged hadrons significantly different energy dependence is observed. 
For comparison, the \EposLong model was employed, known for its alignment with \NASixtyOne results in many cases. Generally, \EposLong indicates similar energy dependence and magnitude of the ratios of cumulants as in measurements performed in $^{40}$Ar+$^{45}$Sc. 
Further studies in Ar+Sc interactions require either increasing the data statistics or finding a more suitable set of quantities that can be utilized in larger centrality bins where volume fluctuations can not be neglected.

\section{Acknowledgments}
We would like to thank the CERN EP, BE, HSE and EN Departments for the
strong support of NA61/SHINE.

This work was supported by
the Hungarian Scientific Research Fund (grant NKFIH 138136\slash137812\slash138152 and TKP2021-NKTA-64),
the Polish Ministry of Science and Higher Education
(DIR\slash WK\slash\-2016\slash 2017\slash\-10-1, WUT ID-UB), the National Science Centre Poland (grants
2014\slash 14\slash E\slash ST2\slash 00018, 
2016\slash 21\slash D\slash ST2\slash 01983, 
2017\slash 25\slash N\slash ST2\slash 02575, 
2018\slash 29\slash N\slash ST2\slash 02595, 
2018\slash 30\slash A\slash ST2\slash 00226, 
2018\slash 31\slash G\slash ST2\slash 03910, 
2020\slash 39\slash O\slash ST2\slash 00277), 
the Norwegian Financial Mechanism 2014--2021 (grant 2019\slash 34\slash H\slash ST2\slash 00585),
the Polish Minister of Education and Science (contract No. 2021\slash WK\slash 10),
the European Union's Horizon 2020 research and innovation programme under grant agreement No. 871072,
the Ministry of Education, Culture, Sports,
Science and Tech\-no\-lo\-gy, Japan, Grant-in-Aid for Sci\-en\-ti\-fic
Research (grants 18071005, 19034011, 19740162, 20740160 and 20039012,22H04943),
the German Research Foundation DFG (grants GA\,1480\slash8-1 and project 426579465),
the Bulgarian Ministry of Education and Science within the National
Roadmap for Research Infrastructures 2020--2027, contract No. D01-374/18.12.2020,
Serbian Ministry of Science, Technological Development and Innovation (grant
OI171002), Swiss Nationalfonds Foundation (grant 200020\-117913/1),
ETH Research Grant TH-01\,07-3, National Science Foundation grant
PHY-2013228 and the Fermi National Accelerator Laboratory (Fermilab),
a U.S. Department of Energy, Office of Science, HEP User Facility
managed by Fermi Research Alliance, LLC (FRA), acting under Contract
No. DE-AC02-07CH11359 and the IN2P3-CNRS (France).\\


\bibliographystyle{na61Utphys}
{\raggedright
\bibliography{na61References}

\providecommand{\href}[2]{#2}\begingroup\raggedright\begin{thebibliography}{10}

\bibitem{Antoniou:2006mh}
N.~Antoniou {\em et~al.}, {[NA61/SHINE} Collab.], ``{Study of hadron production
  in hadron-nucleus and nucleus-nucleus collisions at the CERN SPS},'' 2006.
\newblock {CERN-SPSC-2006-034}.

\bibitem{Abgrall:2014fa}
N.~Abgrall {\em et~al.}, {[NA61/SHINE} Collab.]
  \href{http://dx.doi.org/10.1088/1748-0221/9/06/P06005}{{\em JINST} {\bfseries
  9} (2014) P06005},
\href{http://arxiv.org/abs/1401.4699}{{\ttfamily arXiv:1401.4699
  [physics.ins-det]}}.

\bibitem{Fodor:2004nz}
Z.~Fodor and S.~D. Katz
  \href{http://dx.doi.org/10.1088/1126-6708/2004/04/050}{{\em JHEP} {\bfseries
  0404} (2004) 050},
\href{http://arxiv.org/abs/hep-lat/0402006}{{\ttfamily arXiv:hep-lat/0402006
  [hep-lat]}}.

\bibitem{Stephanov:1999zu}
M.~A. Stephanov, K.~Rajagopal, and E.~V. Shuryak
  \href{http://dx.doi.org/10.1103/PhysRevD.60.114028}{{\em Phys. Rev. D}
  {\bfseries 60} (1999) 114028},
\href{http://arxiv.org/abs/hep-ph/9903292}{{\ttfamily arXiv:hep-ph/9903292
  [hep-ph]}}.

\bibitem{Stephanov:2004xs}
M.~A. Stephanov {\em Acta Phys. Polon. B} {\bfseries 35} (2004) 2939--2962.

\bibitem{Mackowiak-Pawlowska:2021tch}
M.~Ma\'{c}kowiak-Paw{\l}owska, {[NA61/SHINE} Collab.]
  \href{http://dx.doi.org/10.22323/1.380.0238}{{\em PoS} {\bfseries PANIC2021}
  (2022) 238}, \href{http://arxiv.org/abs/2112.01877}{{\ttfamily
  arXiv:2112.01877 [nucl-ex]}}.

\bibitem{ALICE:2021hkc}
S.~Acharya {\em et~al.}, {[ALICE} Collab.]
  \href{http://dx.doi.org/10.1140/epjc/s10052-021-09784-4}{{\em Eur. Phys. J.
  C} {\bfseries 81} no.~11, (2021) 1012},
  \href{http://arxiv.org/abs/2105.05745}{{\ttfamily arXiv:2105.05745
  [nucl-ex]}}.

\bibitem{STAR:2014egu}
L.~Adamczyk {\em et~al.}, {[STAR} Collab.]
  \href{http://dx.doi.org/10.1103/PhysRevLett.113.092301}{{\em Phys. Rev.
  Lett.} {\bfseries 113} (2014) 092301},
  \href{http://arxiv.org/abs/1402.1558}{{\ttfamily arXiv:1402.1558 [nucl-ex]}}.

\bibitem{Asakawa:2015ybt}
M.~Asakawa and M.~Kitazawa
  \href{http://dx.doi.org/10.1016/j.ppnp.2016.04.002}{{\em Prog. Part. Nucl.
  Phys.} {\bfseries 90} (2016) 299--342},
  \href{http://arxiv.org/abs/1512.05038}{{\ttfamily arXiv:1512.05038
  [nucl-th]}}.

\bibitem{Vovchenko:2015pya}
V.~Vovchenko, D.~V. Anchishkin, M.~I. Gorenstein, and R.~V. Poberezhnyuk
  \href{http://dx.doi.org/10.1103/PhysRevC.92.054901}{{\em Phys. Rev. C}
  {\bfseries 92} no.~5, (2015) 054901},
  \href{http://arxiv.org/abs/1506.05763}{{\ttfamily arXiv:1506.05763
  [nucl-th]}}.

\bibitem{Begun:2006jf}
V.~Begun, M.~I. Gorenstein, M.~Hauer, V.~Konchakovski, and O.~Zozulya
  \href{http://dx.doi.org/10.1103/PhysRevC.74.044903}{{\em Phys. Rev. C}
  {\bfseries 74} (2006) 044903},
\href{http://arxiv.org/abs/nucl-th/0606036}{{\ttfamily arXiv:nucl-th/0606036
  [nucl-th]}}.

\bibitem{Bialas:1976ed}
A.~Bia{\l}as, M.~Bleszy\'{n}ski, and W.~Czy\.{z}
\href{http://dx.doi.org/10.1016/0550-3213(76)90329-1}{{\em Nucl.\ Phys. B}
  {\bfseries 111} (1976) 461}.

\bibitem{Begun:2017gsw}
V.~Begun and M.~Mackowiak-Pawlowska
  \href{http://dx.doi.org/10.22323/1.311.0065}{{\em PoS} {\bfseries CPOD2017}
  (2018) 065}, \href{http://arxiv.org/abs/1712.05805}{{\ttfamily
  arXiv:1712.05805 [nucl-th]}}.

\bibitem{Mackowiak-Pawlowska:2021sea}
M.~Mackowiak-Pawlowska, M.~Naskr\k{e}t, and M.~Gazdzicki
  \href{http://dx.doi.org/10.1016/j.nuclphysa.2021.122258}{{\em Nucl. Phys. A}
  {\bfseries 1014} (2021) 122258},
  \href{http://arxiv.org/abs/2102.11186}{{\ttfamily arXiv:2102.11186
  [hep-ex]}}.

\bibitem{Banas:2018sak}
D.~Bana\'{s}, A.~Kubala-Kuku\'{s}, M.~Rybczy\'{n}ski, I.~Stabrawa, and
  G.~Stefanek \href{http://dx.doi.org/10.1140/epjp/i2019-12465-9}{{\em Eur.
  Phys. J. Plus} {\bfseries 134} no.~1, (2019) 44},
  \href{http://arxiv.org/abs/1808.10377}{{\ttfamily arXiv:1808.10377
  [nucl-ex]}}.

\bibitem{Detector_acceptance}
J.~Cybowska, ``Detector acceptance maps for track selection.''
  \url{https://edms.cern.ch/document/2487456/1}, 2020.
\newblock CERN EDMS.

\bibitem{PSD_acceptance}
A.~Seryakov, ``{PSD} acceptance maps for event selection.''
  \url{https://edms.cern.ch/document/1867336/1}, 2017.
\newblock CERN EDMS.

\bibitem{Seryakov:2018kzj}
A.~Seryakov, {[NA61/SHINE} Collab.]
  \href{http://dx.doi.org/10.22323/1.311.0050}{{\em PoS} {\bfseries CPOD2017}
  (2018) 050}.

\bibitem{NA61SHINE:2021nye}
A.~Acharya {\em et~al.}, {[NA61/SHINE} Collab.]
  \href{http://dx.doi.org/10.1140/epjc/s10052-021-09135-3}{{\em Eur. Phys. J.
  C} {\bfseries 81} no.~5, (2021) 397},
  \href{http://arxiv.org/abs/2101.08494}{{\ttfamily arXiv:2101.08494
  [hep-ex]}}.

\bibitem{DAgostini}
G.~D'Agostini \href{http://dx.doi.org/10.1016/0168-9002(95)00274-X}{{\em Nucl.
  Instrum. Meth.} {\bfseries A 362} (1995) 487--498}.

\bibitem{unfolding}
T.~Adye {\em Proceedings of the PHYSTAT 2011 Workshop on Statistical Issues
  Related to Discovery Claims in Search Experiments and Unfolding, CERN,
  Geneva, Switzerland, CERN-2011-006} (2011) 313--318.
  \url{http://hepunx.rl.ac.uk/~adye/software/unfold/RooUnfold.html}.

\bibitem{Werner:2008zza}
K.~Werner
\href{http://dx.doi.org/10.1016/j.nuclphysbps.2007.10.012}{{\em Nucl. Phys.
  Proc. Suppl.} {\bfseries 175-176} (2008) 81--87}.

\bibitem{NA61SHINE:2023kod}
H.~Adhikary {\em et~al.}, {[NA61/SHINE} Collab.]
  \href{http://dx.doi.org/10.1140/epjc/s10052-024-13076-y}{{\em Eur. Phys. J.
  C} {\bfseries 84} no.~9, (2024) 921},
  \href{http://arxiv.org/abs/2312.13706}{{\ttfamily arXiv:2312.13706
  [hep-ex]}}. [Erratum: Eur. Phys. J. C 85, 341 (2025)].

\bibitem{Agostinelli:2002hh}
S.~Agostinelli {\em et~al.}, {[GEANT4} Collab.]
\href{http://dx.doi.org/10.1016/S0168-9002(03)01368-8}{{\em Nucl.\ Instrum.\
  Meth.} {\bfseries A506} (2003) 250}.

\bibitem{moore}
T.~C. Hesterberg, S.~Monaghan, D.~S. Moore, A.~Clipson, and R.~Epstein,
  ``{Bootstrap methods and permutation tests}.''. WHFreeman and Company, New
  York, 2003.

\bibitem{Cybowska:2023PhD}
J. Cybowska, ''Multiplicity and net-electric charge fluctuations in central
  Ar+Sc interactions at SPS energies measured in the NA61/SHINE experiment'',
  PhD thesis, Warsaw University of Technology, 2023.
  \url{https://edms.cern.ch/document/3220743/1}.

\end{thebibliography}\endgroup
}

\newpage
{\Large The \NASixtyOne Collaboration}
\bigskip
\begin{sloppypar}

\noindent
{H.\;Adhikary~\href{https://orcid.org/0000-0002-5746-1268}{\includegraphics[height=1.7ex]{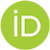}}\textsuperscript{\,11}},
{P.\;Adrich~\href{https://orcid.org/0000-0002-7019-5451}{\includegraphics[height=1.7ex]{orcid-logo.png}}\textsuperscript{\,13}},
{K.K.\;Allison~\href{https://orcid.org/0000-0002-3494-9383}{\includegraphics[height=1.7ex]{orcid-logo.png}}\textsuperscript{\,25}},
{N.\;Amin~\href{https://orcid.org/0009-0004-7572-3817}{\includegraphics[height=1.7ex]{orcid-logo.png}}\textsuperscript{\,4}},
{E.V.\;Andronov~\href{https://orcid.org/0000-0003-0437-9292}{\includegraphics[height=1.7ex]{orcid-logo.png}}\textsuperscript{\,21}},
{I.-C.\;Arsene~\href{https://orcid.org/0000-0003-2316-9565}{\includegraphics[height=1.7ex]{orcid-logo.png}}\textsuperscript{\,10}},
{M.\;Bajda~\href{https://orcid.org/0009-0005-8859-1099}{\includegraphics[height=1.7ex]{orcid-logo.png}}\textsuperscript{\,14}},
{Y.\;Balkova~\href{https://orcid.org/0000-0002-6957-573X}{\includegraphics[height=1.7ex]{orcid-logo.png}}\textsuperscript{\,16}},
{D.\;Battaglia~\href{https://orcid.org/0000-0002-5283-0992}{\includegraphics[height=1.7ex]{orcid-logo.png}}\textsuperscript{\,24}},
{A.\;Bazgir~\href{https://orcid.org/0000-0003-0358-0576}{\includegraphics[height=1.7ex]{orcid-logo.png}}\textsuperscript{\,11}},
{S.\;Bhosale~\href{https://orcid.org/0000-0001-5709-4747}{\includegraphics[height=1.7ex]{orcid-logo.png}}\textsuperscript{\,12}},
{M.\;Bielewicz~\href{https://orcid.org/0000-0001-8267-4874}{\includegraphics[height=1.7ex]{orcid-logo.png}}\textsuperscript{\,13}},
{A.\;Blondel~\href{https://orcid.org/0000-0002-1597-8859}{\includegraphics[height=1.7ex]{orcid-logo.png}}\textsuperscript{\,3}},
{M.\;Bogomilov~\href{https://orcid.org/0000-0001-7738-2041}{\includegraphics[height=1.7ex]{orcid-logo.png}}\textsuperscript{\,2}},
{Y.\;Bondar~\href{https://orcid.org/0000-0003-2773-9668}{\includegraphics[height=1.7ex]{orcid-logo.png}}\textsuperscript{\,11}},
{W.\;Bryli\'nski~\href{https://orcid.org/0000-0002-3457-6601}{\includegraphics[height=1.7ex]{orcid-logo.png}}\textsuperscript{\,19}},
{J.\;Brzychczyk~\href{https://orcid.org/0000-0001-5320-6748}{\includegraphics[height=1.7ex]{orcid-logo.png}}\textsuperscript{\,14}},
{M.\;Buryakov~\href{https://orcid.org/0009-0008-2394-4967}{\includegraphics[height=1.7ex]{orcid-logo.png}}\textsuperscript{\,20}},
{A.F.\;Camino\textsuperscript{\,27}},
{Y.D.\;Chandak~\href{https://orcid.org/0009-0009-2080-566X}{\includegraphics[height=1.7ex]{orcid-logo.png}}\textsuperscript{\,25}},
{M.\;\'Cirkovi\'c~\href{https://orcid.org/0000-0002-4420-9688}{\includegraphics[height=1.7ex]{orcid-logo.png}}\textsuperscript{\,22}},
{M.\;Csan\'ad~\href{https://orcid.org/0000-0002-3154-6925}{\includegraphics[height=1.7ex]{orcid-logo.png}}\textsuperscript{\,6}},
{J.\;Cybowska~\href{https://orcid.org/0000-0003-2568-3664}{\includegraphics[height=1.7ex]{orcid-logo.png}}\textsuperscript{\,19}},
{T.\;Czopowicz~\href{https://orcid.org/0000-0003-1908-2977}{\includegraphics[height=1.7ex]{orcid-logo.png}}\textsuperscript{\,11}},
{C.\;Dalmazzone~\href{https://orcid.org/0000-0001-6945-5845}{\includegraphics[height=1.7ex]{orcid-logo.png}}\textsuperscript{\,3}},
{N.\;Davis~\href{https://orcid.org/0000-0003-3047-6854}{\includegraphics[height=1.7ex]{orcid-logo.png}}\textsuperscript{\,12}},
{A.\;Dmitriev~\href{https://orcid.org/0000-0001-7853-0173}{\includegraphics[height=1.7ex]{orcid-logo.png}}\textsuperscript{\,20}},
{P.~von\;Doetinchem~\href{https://orcid.org/0000-0002-7801-3376}{\includegraphics[height=1.7ex]{orcid-logo.png}}\textsuperscript{\,26}},
{W.\;Dominik~\href{https://orcid.org/0000-0001-7444-9239}{\includegraphics[height=1.7ex]{orcid-logo.png}}\textsuperscript{\,17}},
{J.\;Dumarchez~\href{https://orcid.org/0000-0002-9243-4425}{\includegraphics[height=1.7ex]{orcid-logo.png}}\textsuperscript{\,3}},
{R.\;Engel~\href{https://orcid.org/0000-0003-2924-8889}{\includegraphics[height=1.7ex]{orcid-logo.png}}\textsuperscript{\,4}},
{G.A.\;Feofilov~\href{https://orcid.org/0000-0003-3700-8623}{\includegraphics[height=1.7ex]{orcid-logo.png}}\textsuperscript{\,21}},
{L.\;Fields~\href{https://orcid.org/0000-0001-8281-3686}{\includegraphics[height=1.7ex]{orcid-logo.png}}\textsuperscript{\,24}},
{Z.\;Fodor~\href{https://orcid.org/0000-0003-2519-5687}{\includegraphics[height=1.7ex]{orcid-logo.png}}\textsuperscript{\,5,18}},
{M.\;Friend~\href{https://orcid.org/0000-0003-4660-4670}{\includegraphics[height=1.7ex]{orcid-logo.png}}\textsuperscript{\,7}},
{M.\;Ga\'zdzicki~\href{https://orcid.org/0000-0002-6114-8223}{\includegraphics[height=1.7ex]{orcid-logo.png}}\textsuperscript{\,11}},
{K.E.\;Gollwitzer\textsuperscript{\,23}},
{O.\;Golosov~\href{https://orcid.org/0000-0001-6562-2925}{\includegraphics[height=1.7ex]{orcid-logo.png}}\textsuperscript{\,21}},
{V.\;Golovatyuk~\href{https://orcid.org/0009-0006-5201-0990}{\includegraphics[height=1.7ex]{orcid-logo.png}}\textsuperscript{\,20}},
{M.\;Golubeva~\href{https://orcid.org/0009-0003-4756-2449}{\includegraphics[height=1.7ex]{orcid-logo.png}}\textsuperscript{\,21}},
{K.\;Grebieszkow~\href{https://orcid.org/0000-0002-6754-9554}{\includegraphics[height=1.7ex]{orcid-logo.png}}\textsuperscript{\,19}},
{F.\;Guber~\href{https://orcid.org/0000-0001-8790-3218}{\includegraphics[height=1.7ex]{orcid-logo.png}}\textsuperscript{\,21}},
{P.G.\;Hurh~\href{https://orcid.org/0000-0002-9024-5399}{\includegraphics[height=1.7ex]{orcid-logo.png}}\textsuperscript{\,23}},
{S.\;Ilieva~\href{https://orcid.org/0000-0001-9204-2563}{\includegraphics[height=1.7ex]{orcid-logo.png}}\textsuperscript{\,2}},
{A.\;Ivashkin~\href{https://orcid.org/0000-0003-4595-5866}{\includegraphics[height=1.7ex]{orcid-logo.png}}\textsuperscript{\,21}},
{N.\;Karpushkin~\href{https://orcid.org/0000-0001-5513-9331}{\includegraphics[height=1.7ex]{orcid-logo.png}}\textsuperscript{\,21}},
{M.\;Kie{\l}bowicz~\href{https://orcid.org/0000-0002-4403-9201}{\includegraphics[height=1.7ex]{orcid-logo.png}}\textsuperscript{\,12}},
{V.A.\;Kireyeu~\href{https://orcid.org/0000-0002-5630-9264}{\includegraphics[height=1.7ex]{orcid-logo.png}}\textsuperscript{\,20}},
{R.\;Kolesnikov~\href{https://orcid.org/0009-0006-4224-1058}{\includegraphics[height=1.7ex]{orcid-logo.png}}\textsuperscript{\,20}},
{D.\;Kolev~\href{https://orcid.org/0000-0002-9203-4739}{\includegraphics[height=1.7ex]{orcid-logo.png}}\textsuperscript{\,2}},
{Y.\;Koshio~\href{https://orcid.org/0000-0003-0437-8505}{\includegraphics[height=1.7ex]{orcid-logo.png}}\textsuperscript{\,8}},
{S.\;Kowalski~\href{https://orcid.org/0000-0001-9888-4008}{\includegraphics[height=1.7ex]{orcid-logo.png}}\textsuperscript{\,16}},
{B.\;Koz{\l}owski~\href{https://orcid.org/0000-0001-8442-2320}{\includegraphics[height=1.7ex]{orcid-logo.png}}\textsuperscript{\,19}},
{A.\;Krasnoperov~\href{https://orcid.org/0000-0002-1425-2861}{\includegraphics[height=1.7ex]{orcid-logo.png}}\textsuperscript{\,20}},
{W.\;Kucewicz~\href{https://orcid.org/0000-0002-2073-711X}{\includegraphics[height=1.7ex]{orcid-logo.png}}\textsuperscript{\,15}},
{M.\;Kuchowicz~\href{https://orcid.org/0000-0003-3174-585X}{\includegraphics[height=1.7ex]{orcid-logo.png}}\textsuperscript{\,18}},
{M.\;Kuich~\href{https://orcid.org/0000-0002-6507-8699}{\includegraphics[height=1.7ex]{orcid-logo.png}}\textsuperscript{\,17}},
{A.\;L\'aszl\'o~\href{https://orcid.org/0000-0003-2712-6968}{\includegraphics[height=1.7ex]{orcid-logo.png}}\textsuperscript{\,5}},
{M.\;Lewicki~\href{https://orcid.org/0000-0002-8972-3066}{\includegraphics[height=1.7ex]{orcid-logo.png}}\textsuperscript{\,12}},
{G.\;Lykasov~\href{https://orcid.org/0000-0002-1544-6959}{\includegraphics[height=1.7ex]{orcid-logo.png}}\textsuperscript{\,20}},
{J.R.\;Lyon~\href{https://orcid.org/0009-0003-2579-8821}{\includegraphics[height=1.7ex]{orcid-logo.png}}\textsuperscript{\,26}},
{V.V.\;Lyubushkin~\href{https://orcid.org/0000-0003-0136-233X}{\includegraphics[height=1.7ex]{orcid-logo.png}}\textsuperscript{\,20}},
{M.\;Ma\'ckowiak-Paw{\l}owska~\href{https://orcid.org/0000-0003-3954-6329}{\includegraphics[height=1.7ex]{orcid-logo.png}}\textsuperscript{\,19}},
{B.\;Maksiak~\href{https://orcid.org/0000-0002-7950-2307}{\includegraphics[height=1.7ex]{orcid-logo.png}}\textsuperscript{\,13}},
{A.I.\;Malakhov~\href{https://orcid.org/0000-0001-8569-8409}{\includegraphics[height=1.7ex]{orcid-logo.png}}\textsuperscript{\,20}},
{A.\;Marcinek~\href{https://orcid.org/0000-0001-9922-743X}{\includegraphics[height=1.7ex]{orcid-logo.png}}\textsuperscript{\,12}},
{A.D.\;Marino~\href{https://orcid.org/0000-0002-1709-538X}{\includegraphics[height=1.7ex]{orcid-logo.png}}\textsuperscript{\,25}},
{H.-J.\;Mathes~\href{https://orcid.org/0000-0002-0680-040X}{\includegraphics[height=1.7ex]{orcid-logo.png}}\textsuperscript{\,4}},
{T.\;Matulewicz~\href{https://orcid.org/0000-0003-2098-1216}{\includegraphics[height=1.7ex]{orcid-logo.png}}\textsuperscript{\,17}},
{V.\;Matveev~\href{https://orcid.org/0000-0002-2745-5908}{\includegraphics[height=1.7ex]{orcid-logo.png}}\textsuperscript{\,20}},
{G.L.\;Melkumov~\href{https://orcid.org/0009-0004-2074-6755}{\includegraphics[height=1.7ex]{orcid-logo.png}}\textsuperscript{\,20}},
{A.\;Merzlaya~\href{https://orcid.org/0000-0002-6553-2783}{\includegraphics[height=1.7ex]{orcid-logo.png}}\textsuperscript{\,10}},
{{\L}.\;Mik~\href{https://orcid.org/0000-0003-2712-6861}{\includegraphics[height=1.7ex]{orcid-logo.png}}\textsuperscript{\,15}},
{S.\;Morozov~\href{https://orcid.org/0000-0002-6748-7277}{\includegraphics[height=1.7ex]{orcid-logo.png}}\textsuperscript{\,21}},
{Y.\;Nagai~\href{https://orcid.org/0000-0002-1792-5005}{\includegraphics[height=1.7ex]{orcid-logo.png}}\textsuperscript{\,6}},
{T.\;Nakadaira~\href{https://orcid.org/0000-0003-4327-7598}{\includegraphics[height=1.7ex]{orcid-logo.png}}\textsuperscript{\,7}},
{S.\;Nishimori~\href{https://orcid.org/~0000-0002-1820-0938}{\includegraphics[height=1.7ex]{orcid-logo.png}}\textsuperscript{\,7}},
{A.\;Olivier~\href{https://orcid.org/0000-0003-4261-8303}{\includegraphics[height=1.7ex]{orcid-logo.png}}\textsuperscript{\,24}},
{V.\;Ozvenchuk~\href{https://orcid.org/0000-0002-7821-7109}{\includegraphics[height=1.7ex]{orcid-logo.png}}\textsuperscript{\,12}},
{O.\;Panova~\href{https://orcid.org/0000-0001-5039-7788}{\includegraphics[height=1.7ex]{orcid-logo.png}}\textsuperscript{\,11}},
{V.\;Paolone~\href{https://orcid.org/0000-0003-2162-0957}{\includegraphics[height=1.7ex]{orcid-logo.png}}\textsuperscript{\,27}},
{I.\;Pidhurskyi~\href{https://orcid.org/0000-0001-9916-9436}{\includegraphics[height=1.7ex]{orcid-logo.png}}\textsuperscript{\,11}},
{R.\;P{\l}aneta~\href{https://orcid.org/0000-0001-8007-8577}{\includegraphics[height=1.7ex]{orcid-logo.png}}\textsuperscript{\,14}},
{P.\;Podlaski~\href{https://orcid.org/0000-0002-0232-9841}{\includegraphics[height=1.7ex]{orcid-logo.png}}\textsuperscript{\,17}},
{B.A.\;Popov~\href{https://orcid.org/0000-0001-5416-9301}{\includegraphics[height=1.7ex]{orcid-logo.png}}\textsuperscript{\,20,3}},
{B.\;P\'orfy~\href{https://orcid.org/0000-0001-5724-9737}{\includegraphics[height=1.7ex]{orcid-logo.png}}\textsuperscript{\,5,6}},
{D.S.\;Prokhorova~\href{https://orcid.org/0000-0003-3726-9196}{\includegraphics[height=1.7ex]{orcid-logo.png}}\textsuperscript{\,21}},
{D.\;Pszczel~\href{https://orcid.org/0000-0002-4697-6688}{\includegraphics[height=1.7ex]{orcid-logo.png}}\textsuperscript{\,13}},
{S.\;Pu{\l}awski~\href{https://orcid.org/0000-0003-1982-2787}{\includegraphics[height=1.7ex]{orcid-logo.png}}\textsuperscript{\,16}},
{L.\;Ren~\href{https://orcid.org/0000-0003-1709-7673}{\includegraphics[height=1.7ex]{orcid-logo.png}}\textsuperscript{\,25}},
{V.Z.\;Reyna~Ortiz~\href{https://orcid.org/0000-0002-7026-8198}{\includegraphics[height=1.7ex]{orcid-logo.png}}\textsuperscript{\,11}},
{D.\;R\"ohrich\textsuperscript{\,9}},
{M.\;Roth~\href{https://orcid.org/0000-0003-1281-4477}{\includegraphics[height=1.7ex]{orcid-logo.png}}\textsuperscript{\,4}},
{{\L}.\;Rozp{\l}ochowski~\href{https://orcid.org/0000-0003-3680-6738}{\includegraphics[height=1.7ex]{orcid-logo.png}}\textsuperscript{\,12}},
{M.\;Rumyantsev~\href{https://orcid.org/0000-0001-8233-2030}{\includegraphics[height=1.7ex]{orcid-logo.png}}\textsuperscript{\,20}},
{A.\;Rustamov~\href{https://orcid.org/0000-0001-8678-6400}{\includegraphics[height=1.7ex]{orcid-logo.png}}\textsuperscript{\,1}},
{M.\;Rybczynski~\href{https://orcid.org/0000-0002-3638-3766}{\includegraphics[height=1.7ex]{orcid-logo.png}}\textsuperscript{\,11}},
{A.\;Rybicki~\href{https://orcid.org/0000-0003-3076-0505}{\includegraphics[height=1.7ex]{orcid-logo.png}}\textsuperscript{\,12}},
{D.\;Rybka~\href{https://orcid.org/0000-0002-9924-6398}{\includegraphics[height=1.7ex]{orcid-logo.png}}\textsuperscript{\,13}},
{K.\;Sakashita~\href{https://orcid.org/0000-0003-2602-7837}{\includegraphics[height=1.7ex]{orcid-logo.png}}\textsuperscript{\,7}},
{K.\;Schmidt~\href{https://orcid.org/0000-0002-0903-5790}{\includegraphics[height=1.7ex]{orcid-logo.png}}\textsuperscript{\,16}},
{P.\;Seyboth~\href{https://orcid.org/0000-0002-4821-6105}{\includegraphics[height=1.7ex]{orcid-logo.png}}\textsuperscript{\,11}},
{U.A.\;Shah~\href{https://orcid.org/0000-0002-9315-1304}{\includegraphics[height=1.7ex]{orcid-logo.png}}\textsuperscript{\,11}},
{Y.\;Shiraishi~\href{https://orcid.org/0000-0002-0132-3923}{\includegraphics[height=1.7ex]{orcid-logo.png}}\textsuperscript{\,8}},
{A.\;Shukla~\href{https://orcid.org/0000-0003-3839-7229}{\includegraphics[height=1.7ex]{orcid-logo.png}}\textsuperscript{\,26}},
{M.\;S{\l}odkowski~\href{https://orcid.org/0000-0003-0463-2753}{\includegraphics[height=1.7ex]{orcid-logo.png}}\textsuperscript{\,19}},
{P.\;Staszel~\href{https://orcid.org/0000-0003-4002-1626}{\includegraphics[height=1.7ex]{orcid-logo.png}}\textsuperscript{\,14}},
{G.\;Stefanek~\href{https://orcid.org/0000-0001-6656-9177}{\includegraphics[height=1.7ex]{orcid-logo.png}}\textsuperscript{\,11}},
{J.\;Stepaniak~\href{https://orcid.org/0000-0003-2064-9870}{\includegraphics[height=1.7ex]{orcid-logo.png}}\textsuperscript{\,13}},
{{\L}.\;\'Swiderski~\href{https://orcid.org/0000-0001-5857-2085}{\includegraphics[height=1.7ex]{orcid-logo.png}}\textsuperscript{\,13}},
{J.\;Szewi\'nski~\href{https://orcid.org/0000-0003-2981-9303}{\includegraphics[height=1.7ex]{orcid-logo.png}}\textsuperscript{\,13}},
{R.\;Szukiewicz~\href{https://orcid.org/0000-0002-1291-4040}{\includegraphics[height=1.7ex]{orcid-logo.png}}\textsuperscript{\,18}},
{A.\;Taranenko~\href{https://orcid.org/0000-0003-1737-4474}{\includegraphics[height=1.7ex]{orcid-logo.png}}\textsuperscript{\,21}},
{A.\;Tefelska~\href{https://orcid.org/0000-0002-6069-4273}{\includegraphics[height=1.7ex]{orcid-logo.png}}\textsuperscript{\,19}},
{D.\;Tefelski~\href{https://orcid.org/0000-0003-0802-2290}{\includegraphics[height=1.7ex]{orcid-logo.png}}\textsuperscript{\,19}},
{V.\;Tereshchenko\textsuperscript{\,20}},
{R.\;Tsenov~\href{https://orcid.org/0000-0002-1330-8640}{\includegraphics[height=1.7ex]{orcid-logo.png}}\textsuperscript{\,2}},
{L.\;Turko~\href{https://orcid.org/0000-0002-5474-8650}{\includegraphics[height=1.7ex]{orcid-logo.png}}\textsuperscript{\,18}},
{T.S.\;Tveter~\href{https://orcid.org/0009-0003-7140-8644}{\includegraphics[height=1.7ex]{orcid-logo.png}}\textsuperscript{\,10}},
{M.\;Unger~\href{https://orcid.org/0000-0002-7651-0272~}{\includegraphics[height=1.7ex]{orcid-logo.png}}\textsuperscript{\,4}},
{M.\;Urbaniak~\href{https://orcid.org/0000-0002-9768-030X}{\includegraphics[height=1.7ex]{orcid-logo.png}}\textsuperscript{\,16}},
{D.\;Veberi\v{c}~\href{https://orcid.org/0000-0003-2683-1526}{\includegraphics[height=1.7ex]{orcid-logo.png}}\textsuperscript{\,4}},
{O.\;Vitiuk~\href{https://orcid.org/0000-0002-9744-3937}{\includegraphics[height=1.7ex]{orcid-logo.png}}\textsuperscript{\,18}},
{A.\;Wickremasinghe~\href{https://orcid.org/0000-0002-5325-0455}{\includegraphics[height=1.7ex]{orcid-logo.png}}\textsuperscript{\,23}},
{K.\;Witek~\href{https://orcid.org/0009-0004-6699-1895}{\includegraphics[height=1.7ex]{orcid-logo.png}}\textsuperscript{\,15}},
{K.\;W\'ojcik~\href{https://orcid.org/0000-0002-8315-9281}{\includegraphics[height=1.7ex]{orcid-logo.png}}\textsuperscript{\,16}},
{O.\;Wyszy\'nski~\href{https://orcid.org/0000-0002-6652-0450}{\includegraphics[height=1.7ex]{orcid-logo.png}}\textsuperscript{\,11}},
{A.\;Zaitsev~\href{https://orcid.org/0000-0003-4711-9925}{\includegraphics[height=1.7ex]{orcid-logo.png}}\textsuperscript{\,20}},
{E.\;Zherebtsova~\href{https://orcid.org/0000-0002-1364-0969}{\includegraphics[height=1.7ex]{orcid-logo.png}}\textsuperscript{\,18}},
{E.D.\;Zimmerman~\href{https://orcid.org/0000-0002-6394-6659}{\includegraphics[height=1.7ex]{orcid-logo.png}}\textsuperscript{\,25}}, and
{A.\;Zviagina~\href{https://orcid.org/0009-0007-5211-6493}{\includegraphics[height=1.7ex]{orcid-logo.png}}\textsuperscript{\,21}}

\end{sloppypar}

\noindent
\textsuperscript{1}~National Nuclear Research Center, Baku, Azerbaijan\\
\textsuperscript{2}~Faculty of Physics, University of Sofia, Sofia, Bulgaria\\
\textsuperscript{3}~LPNHE, Sorbonne University, CNRS/IN2P3, Paris, France\\
\textsuperscript{4}~Karlsruhe Institute of Technology, Karlsruhe, Germany\\
\textsuperscript{5}~HUN-REN Wigner Research Centre for Physics, Budapest, Hungary\\
\textsuperscript{6}~E\"otv\"os Lor\'and University, Budapest, Hungary\\
\textsuperscript{7}~Institute for Particle and Nuclear Studies, Tsukuba, Japan\\
\textsuperscript{8}~Okayama University, Japan\\
\textsuperscript{9}~University of Bergen, Bergen, Norway\\
\textsuperscript{10}~University of Oslo, Oslo, Norway\\
\textsuperscript{11}~Jan Kochanowski University, Kielce, Poland\\
\textsuperscript{12}~Institute of Nuclear Physics, Polish Academy of Sciences, Cracow, Poland\\
\textsuperscript{13}~National Centre for Nuclear Research, Warsaw, Poland\\
\textsuperscript{14}~Jagiellonian University, Cracow, Poland\\
\textsuperscript{15}~AGH - University of Krakow, Poland\\
\textsuperscript{16}~University of Silesia, Katowice, Poland\\
\textsuperscript{17}~University of Warsaw, Warsaw, Poland\\
\textsuperscript{18}~University of Wroc{\l}aw,  Wroc{\l}aw, Poland\\
\textsuperscript{19}~Warsaw University of Technology, Warsaw, Poland\\
\textsuperscript{20}~Joint Institute for Nuclear Research, Dubna, International Organization\\
\textsuperscript{21}~Affiliated with an institution formerly covered by a cooperation agreement with CERN\\
\textsuperscript{22}~University of Belgrade, Belgrade, Serbia\\
\textsuperscript{23}~Fermilab, Batavia, USA\\
\textsuperscript{24}~University of Notre Dame, Notre Dame, USA\\
\textsuperscript{25}~University of Colorado, Boulder, USA\\
\textsuperscript{26}~University of Hawaii at Manoa, Honolulu, USA\\
\textsuperscript{27}~University of Pittsburgh, Pittsburgh, USA\\

\end{document}